\pdfoutput=1
\documentclass[11pt]{JHEP3}
\input epsf.tex
\epsfclipon

\usepackage{color}
\let\normalcolor\relax
\usepackage{epsfig}
\usepackage{epstopdf}
\usepackage{epsf}
\input{epsf.sty}
\usepackage{graphicx,amsmath,amssymb}
\usepackage{epsfig,multicol}
\usepackage{multirow}
\allowdisplaybreaks

\usepackage{bbm,bm,amsmath,amssymb}
 
\usepackage[normalem]{ulem}

\usepackage{comment}
\def\be{\begin{equation}}
\def\ee{\end{equation}}
\def\figs/B{B}
\def\bea{\begin{eqnarray}}
\def\eea{\end{eqnarray}}
\def\bg{\begin{eqnarray}}
\def\nd{\end{eqnarray}}

\title{How a four-dimensional de Sitter solution remains outside the swampland}
\author{Keshav Dasgupta$^{1}$, Maxim Emelin$^{1}$, Mir Mehedi Faruk$^{1}$,  
	Radu Tatar$^{2}$\\
	\vskip.03in
	${}^1$ Department of Physics, McGill University, Montr\'{e}al, Qu\'{e}bec, H3A 2T8, Canada \\
	${}^2$ Department of Mathematical Sciences,
University of Liverpool,  Liverpool, L69 7ZL, United Kingdom \\
	{\tt keshav@hep.physics.mcgill.ca, maxim.emelin@mail.mcgill.ca}
	~~{\tt  mir.faruk@mail.mcgill.ca, Radu.Tatar@Liverpool.ac.uk}}

\date{\today}
\maketitle

\abstract{We argue that, in the presence of time-dependent fluxes and quantum corrections, four-dimensional de Sitter solutions should appear in the type IIB string landscape and not in the swampland. Our construction considers generic choices of local and non-local quantum terms and satisfies the no-go and the swampland criteria, the latter being recently upgraded using the trans-Planckian cosmic censorship. Interestingly, both time-independent Newton constant and moduli stabilization may be achieved in such backgrounds even in the presence of time-dependent fluxes and internal spaces. However, once the time-dependence is switched off, any four-dimensional solution with de Sitter isometries appears to have no simple effective field theory descriptions and is back in the swampland.}

\begin{document}

\hskip3.4in {\it Are you watching closely?}

\hskip3.2in $-$ Christopher Nolan's ``The Prestige"

\section{Introduction \label{intro}}

The late-time behavior of our universe appears to be that of accelerated expansion. De Sitter space is the maximally symmetric space with this property and it is therefore natural to attempt its construction within string theory. Several scenarios by which this may be achieved have been proposed \cite{KKLT}, but a completely top-down construction is still lacking. Furthermore, various technical issues within these constructions \cite{bena, sav, westphal} (see however \cite{soler} and \cite{kalloshevan} for possible ways to reconcile some of the issues), combined with more general arguments about the nature of effective theories compatible with quantum gravity \cite{vafa} have cast some doubt about the possibility of realizing meta-stable de Sitter solutions, suggesting instead that string theory prefers quintessence models, the latter however having its own share of problems \cite{HB}.

In previous work \cite{nogo, nodS, deft}, we have investigated the possibility of constructing type IIB de Sitter solutions by analyzing the M-theory uplift of a de Sitter ansatz. Investigations of quantum corrections to the equations of motion revealed potential problems for the existence of an effective theory description of the de Sitter ansatze, when all the internal degrees of freedom were taken to be time-independent. On the other hand, when time-dependence is introduced, these problems go away and appear to permit time-dependent de Sitter compactifications. Interestingly, the space-time constructed in \cite{deft} is {\it exactly} de Sitter in the space-time directions with all time-dependence being confined to the internal space, but in such a way as to preserve Newton's constant. One of the goal of our work is to elaborate on the  precise sources of the breakdown of effective theory in the time-independent cases that were not studied in \cite{deft}. Another goal is to elaborate upon the most generic sources of local and non-local quantum terms and argue from there how a four-dimensional de Sitter solution might be constructed.

In section \ref{qotom1}, we analyze the set-up and compare and contrast the time-independent and time-dependent de Sitter ansatzes. We elaborate on the form of the higher derivative corrections which involve localized fluxes and show exactly how they may lead to a breakdown of effective theory, by requiring that an infinite number of operators be switched on, with no relative suppression between them. We also show how the time-dependent ansatze avoids these difficulties.

In section \ref{qotom2}, we discuss the non-local corrections, which were also considered in the previous work. Here however we elaborate on a more generic set-up. In the time-dependent case, they too present no problems for the construction, however in the time-independent case they risk presenting additional problems for the effective theory over and above those given in section \ref{qotom1}. The non-local terms must go away in the known limits of non-compact 11-dimensional supergravity and type IIB compactifications to flat space. This can happen as consequence of the low-energy limit or through various dependencies on the volumes of cycles, or a combination thereof. We classify the possible non-local corrections according to their mechanism in these limits. We show that certain subsets of these corrections may break the effective theory, by introducing infinitely many degrees of freedom into the theory. Again, we show how the time-dependent ansatze avoids these issues.

In section \ref{later}, we discuss how a late time de Sitter solution satisfies both the no-go and the swampland criteria. We extend the results of \cite{deft} in a broader framework in light of the results of the previous sections. We also provide brief discussions on how moduli stabilization may be addressed in a time-dependent scenario. We conclude with a summary and discussion of our results.

\section{Analysis with local quantum terms \label{qotom1}}

In a recent paper \cite{deft}, we argued how one should be able to get a four-dimensional de Sitter space in the type IIB string landscape once certain time-dependences are switched on. In this paper we want to elaborate more on how the local and non-local quantum terms conspire to help us realize four-dimensional de Sitter solutions in the IIB landscape. We start by first categorizing two possible backgrounds in IIB. The first one may be written as:

{\footnotesize
\bg\label{pyncmey1}
ds^2=\frac{1}{\Lambda(t)H^2}(-dt^2+dx_1^2
+dx_2^2+dx_3^2)+ H^2 \Big({g}_{\alpha\beta}(y)dy^\alpha dy^\beta
+ {g}_{mn}(y)dy^m dy^n\Big), \nd}
where $H^2(y)$ is the warp-factor; and 
we have expressed the compact six-dimensional internal space as ${\cal M}_4 \times {\cal M}_2$ with 
${\cal M}_2$ being parametrized by coordinates $y^\alpha$ and ${\cal M}_4$ being parametrized by 
coordinates $y^m$. The above metric configuration would require background fluxes, such as three-form and five-form fluxes, along-with possible axio-dilaton. We will assume that the three and the five-form fluxes are all time-independent, the axion vanishes, and the dilaton (or the IIB coupling constant) is a constant. Such a background may be compared to another possible background in IIB, which may be expressed as:

{\footnotesize
\bg\label{pyncmey2}
ds^2=\frac{1}{\Lambda(t)H^2}(-dt^2+dx_1^2
+dx_2^2+dx_3^2)+ H^2 \Big(F_1(t){g}_{\alpha\beta}(y)dy^\alpha dy^\beta
+ F_2(t){g}_{mn}(y)dy^m dy^n\Big), \nd}
where we see that it differs from \eqref{pyncmey1} by the presence of time-dependent warp-factors $F_1(t)$ and $F_2(t)$. However now we will also switch on time-dependent three and five-form fluxes, but keep vanishing axion and constant dilaton. The question that we ask is:  which of the two backgrounds \eqref{pyncmey1} and \eqref{pyncmey2}, along with their corresponding flux components, solve the IIB equations of motion?

The answer that we gave in \cite{deft} is that the second background \eqref{pyncmey2} can actually solve the
IIB equations of motion, but only in the presence of the local and non-local quantum corrections.  The first 
background, \eqref{pyncmey1}, with time-independent internal space, can {\it never} solve the IIB equations of motion no matter how many quantum corrections are poured in! In fact there is no simple EFT description possible with a background with four-dimensional de Sitter isometries and time-independent background fluxes. Thus 
\eqref{pyncmey1} is truly in the {\it swampland} \cite{vafa} as shown in \cite{nogo, nodS}.

The next question then is, why does the background \eqref{pyncmey2} solve the quantum corrected equations of motion whereas the  background \eqref{pyncmey1} fails to do so?
This is one aspect that we want to elaborate here, in particular the role that the local and the non-local quantum terms play in realizing such a background in the IIB landscape.

The analysis of the quantum terms, and the equations of motion, gets much easier by lifting the IIB background to M-theory, as shown in \cite{deft}. In M-theory, all we need is the metric and the G-flux components that form the main ingredients in describing the {\it local} quantum terms. The quantum series may be expanded in powers of the type IIA string coupling, $g_s$,  as well as in powers of $M_p$. The IIA string coupling, which now becomes a time-dependent function, can be written 
as\footnote{In this language $\Lambda \vert t\vert^2 \to 0$ or $g_s \to 0$ is late time in the sense that 
$\Lambda \vert t\vert^2 \to \infty$ is the big-bang time and $\Lambda \vert t\vert^2 \to 1$ is the inflationary time. They are respectively related to $g_s \to \infty$ and $g_s \to 1$. \label{godfather}} :
\bg\label{elkeS}
\left({g_s\over H}\right)^2 = \Lambda(t) =  \Lambda \vert t \vert^2, \nd
where $H(y)$ is the same warp-factor appearing in \eqref{pyncmey2}. We also want the four-dimensional Newton's constant to be time independent, so we take $F_i(t)$ to satisfy $F_1(t) F^2_2(t) = 1$ in their $g_s$ expansions:
\bg\label{bobby2}
F_2(t) & = & \sum_{k, n \ge  0} C_{kn} \left({g_s\over H}\right)^{2k/3} 
{\rm exp}\left(-{~n H^{1/3} \over g^{1/3}_s}\right) \cr
F_1(t) & = & F_2^{-2}(t) =   \sum_{k, n >  0} \widetilde{C}_{kn} 
\left({g_s\over H}\right)^{2k/3} {\rm exp}\left(-{~n H^{1/3} \over g^{1/3}_s}\right), \nd
where note that we have expressed the series in terms of powers of $g_s$ and 
${\rm exp}\left(-{1\over g^{1/3}_s}\right)$.
The latter appears from resurgent sum of the inverse $g_s$ powers and should take the above form if the series 
in \eqref{bobby2} have to make sense. The appearance of $1/3$ powers have been explained in details in \cite{deft}, and has to do how the background equations of motion are written in M-theory. 

The advantage of expressing the inverse $g_s$ series in terms of exponential series is two-fold. One, is of course the resurgent sum that collects all the series that are non-perturbative in $g_s$ as a series in powers of ${\rm exp}\left(-{1\over g^{1/3}_s}\right)$. This is useful because at late time, where $g_s \to 0$, the exponential factor decouples and therefore we can only have series perturbative in $g_s$. The late time is also useful because we expect four-dimensional de Sitter solutions to appear at late time in the cosmological evolution. Two, the exponential series provides a hint that the other fields may also be expressed in the same vein. For example we can express the G-flux components as the following series:   
\bg\label{frostgiant2}
{\bf G}_{MNPQ}(y, t) = \sum_{k, n \ge 0} {\cal G}^{(k, n)}_{MNPQ}(y) 
\left({g_s\over H}\right)^{2k/3}
{\rm exp}\left(-{~n H^{1/3} \over g^{1/3}_s}\right), \nd
where again at late time, the non-perturbative piece in \eqref{frostgiant2} decouples. Taking this into account we can effectively keep $n = 0$ in \eqref{frostgiant2}, but $k$ is allowed to take any lower limit. In \cite{deft} we showed that taking $k \ge {3\over 2}$ ($k \in {\mathbb{Z}\over 2}$) we can allow consistent solution to exist in the presence of quantum corrections, although smaller bound for $k$ is also possible. For example all G-flux components of the form ${\bf G}_{MNPQ}$, except the G-flux components ${\bf G}_{MNab}$, can have $k \ge 0$. Here $y^M$ denotes the coordinates of the eight-dimensional internal space ${\cal M}_4 \times {\cal M}_2 \times 
{\mathbb{T}^2\over {\cal G}}$.
The remaining three set of components ${\bf G}_{MNab}$ are bounded from below by $k \ge {3\over 2}$. 

All the G-flux components, curvature tensors and various derivatives acting on them may be used to express the {\it local} quantum terms. The local quantum terms may be expressed in powers of $g_s$, forming the perturbative series, and in powers of ${\rm exp}\left(-{1\over g^{1/3}_s}\right)$, forming the non-perturbative series. The latter coming from the wrapped branes, instantons etc. Again, in the limit $g_s \to 0$, the non-perturbative series will decouple and the local quantum terms may be expressed as a perturbative series in $g_s$. We can then express the local quantum terms in the following way \cite{deft}:
\bg\label{phingsha0} 
\mathbb{Q}_{\rm T}^{(\{l_i\}, n)} & = & {\bf g}^{m_i m'_i}{\bf g}^{m_l m'_l}.... {\bf g}^{j_k j'_k} 
\partial_{m_1}..\partial_{m_{n_1}}\partial_{\alpha_1}..\partial_{\alpha_{n_2}}
\left({\bf R}_{mnpq}\right)^{l_1} \left({\bf R}_{abab}\right)^{l_2}\left({\bf R}_{pqab}\right)^{l_3}\left({\bf R}_{\alpha a b \beta}\right)^{l_4} \nonumber\\
&\times& \left({\bf R}_{\alpha\beta mn}\right)^{l_5}\left({\bf R}_{\alpha\beta\alpha\beta}\right)^{l_6}
\left({\bf R}_{ijij}\right)^{l_7}\left({\bf R}_{ijmn}\right)^{l_8}\left({\bf R}_{iajb}\right)^{l_9}
\left({\bf R}_{i\alpha j \beta}\right)^{l_{10}}\left({\bf R}_{0mnp}\right)^{l_{11}}
\nonumber\\
& \times & \left({\bf R}_{0m0n}\right)^{l_{12}}\left({\bf R}_{0i0j}\right)^{l_{13}}\left({\bf R}_{0a0b}\right)^{l_{14}}\left({\bf R}_{0\alpha 0\beta}\right)^{l_{15}}
\left({\bf R}_{0\alpha\beta m}\right)^{l_{16}}\left({\bf R}_{0abm}\right)^{l_{17}}\left({\bf R}_{0ijm}\right)^{l_{18}}
\nonumber\\
& \times & \left({\bf R}_{mnp\alpha}\right)^{l_{19}}\left({\bf R}_{m\alpha ab}\right)^{l_{20}}
\left({\bf R}_{m\alpha\alpha\beta}\right)^{l_{21}}\left({\bf R}_{m\alpha ij}\right)^{l_{22}}
\left({\bf R}_{0mn \alpha}\right)^{l_{23}}\left({\bf R}_{0m0\alpha}\right)^{l_{24}}
\left({\bf R}_{0\alpha\beta\alpha}\right)^{l_{25}}
\nonumber\\
&\times& \left({\bf R}_{0ab \alpha}\right)^{l_{26}}\left({\bf R}_{0ij\alpha}\right)^{l_{27}}
\left({\bf G}_{mnpq}\right)^{l_{28}}\left({\bf G}_{mnp\alpha}\right)^{l_{29}}
\left({\bf G}_{mnpa}\right)^{l_{30}}\left({\bf G}_{mn\alpha\beta}\right)^{l_{31}}
\left({\bf G}_{mn\alpha a}\right)^{l_{32}}\nonumber\\
&\times&\left({\bf G}_{m\alpha\beta a}\right)^{l_{33}}\left({\bf G}_{0ijm}\right)^{l_{34}} 
\left({\bf G}_{0ij\alpha}\right)^{l_{35}}
\left({\bf G}_{mnab}\right)^{l_{36}}\left({\bf G}_{ab\alpha\beta}\right)^{l_{37}}
\left({\bf G}_{m\alpha ab}\right)^{l_{38}},
\nd
where we have used bold-face for the metric and the curvature components to denote the usage of warped eleven-dimensional metric components. The raising of the curvature and the G-flux components using the $l_i$ exponents with $n \equiv (n_1, n_2)$ denoting the derivative actions along ${\cal M}_4$ and 
${\cal M}_2$ respectively, have been defined in \cite{deft} which the readers may look up for more details.  The quantum terms of the form \eqref{phingsha0} contribute to the three-dimensional potential in the following way\footnote{We are ignoring constant coefficients that in-turn should be specified from our choice of the theory.}:
\bg\label{ducksoup2}
 \mathbb{V}_Q \equiv \sum_{\{l_i\}, n}
 \int d^8 y \sqrt{{\bf g}_8} \left({\mathbb{Q}_T^{(\{l_i\}, n)} \over M_p^{\sigma(\{l_i\}, n)-8}}\right),
 \nd
which precisely captures the complete contributions from the local perturbative quantum terms. Note that 
\eqref{ducksoup2} is the most generic answer we can have here for the late time physics. It should be clear that:
\bg\label{mcgillmey2}
\left(\mathbb{Q}_{\rm T}^{(\{l_i\}, n)}\right) \otimes  \left(\mathbb{Q}_{\rm T}^{(\{l_j\}, m)}\right) 
  \equiv \mathbb{Q}_{\rm T}^{(\{l_i + l_j\}, n+ m)}, \nd
which provides a semi-group structure to the quantum terms \eqref{phingsha0}. This is of course a consequence of the underlying renormalization group structure to the quantum terms, which arises from integrating out those components of fluxes, metric and curvatures that would potentially ruin the de Sitter isometries in the IIB side. The procedure of  integrating out modes requires adding a small mass to those components and the usage of path integral formalism to perform the required operation.
An example was provided in \cite{deft} to show how this influences the dynamics of the system.   

\subsection{Various realizations of de Sitter vacua from M-theory \label{twinpeaks}}

The above procedure is a bit more subtle than what one would expect from our discussion. Question is what modes are we integrating out? In the standard Wilsonian way of doing it, which goes for the time-independent backgrounds, we integrate out all the high energy modes and write an effective action in lower dimensions. The low energy modes, corresponding to fields that would potentially ruin the four-dimensional de Sitter isometries, are given small masses to facilitate the integration procedure. The situation however becomes more subtle when we allow inherent time dependences to the background fields because the modes themselves could acquire additional time-dependences and therefore could in principle get red-shifted. How does the integration procedure works now? In the following, let us discuss two possible scenarios associated to this.

\vskip.1in

\noindent {\it Case 1: de Sitter solution as a background non-supersymmetric vacuum solution}

\vskip.1in

\noindent In the first case, let us assume that the de Sitter solution in \eqref{pyncmey2}, along-with all the time-dependent fluxes,  appears as the background {\it vacuum} solution in IIB. As pointed out in \cite{deft}, there are a few issues with such a viewpoint, which may also be seen from its M-theory uplift. The M-theory background, which appears as eq (3.3) in \cite{deft}, is non-supersymmetric and therefore viewing it as a background solution brings back the un-cancelled vacuum energies from all the modes in the theory. This leads to the vacuum energy problem whose renormalizability, as far as we know, is an unsolved issue.
In addition to that, we expect that, when stretched back in time, any modes in the theory, will necessarily take us to the UV or the trans-Planckian regime. This is the so-called Trans-Planckian problem in cosmology \cite{transplanckian}, and might appear here too despite the fact that we don't have access to $g_s \ge 1$ regime\footnote{A more subtle question is whether the far UV modes have well defined dynamics in string theory when we choose a de Sitter vacuum, i.e whether the UV dynamics lead to unitary interactions over a de Sitter vacuum or not. If the latter is true, then these modes shouldn't be allowed to appear at low energies thus leading to the trans-Planckian censorship conjecture (TCC). \label{tcc}}. For the four-dimensional de Sitter solution this issue was addressed in 
\cite{danielsson2}
to motivate the quantum swampland, and more recently in \cite{vafa2}, to motivate the original swampland criteria \cite{vafa}.  See also the interesting recent work \cite{brahma} where the trans-Planckian censorship conjecture is derived from the swampland distance conjecture.

The issue with such a viewpoint is that a simple Wilsonian effective action cannot be easily defined. For example once we write an effective action by integrating out certain set of UV modes, say for example we integrate out $k \ge k_0$ modes, these modes would re-appear as red-shifted IR modes at late time, thus ruining the entire procedure.  One could however say that since the $k < k_0$ modes  would also be red-shifted, say for example from $k_0$ to $k_1$ where $k_1 < k_0$, then the {\it depleted} region between 
$k_0$ to $k_1$ gets filled up by the red-shifted UV modes. This way the Wilsonian effective action may not change at all. This may be an intriguing way out, but it depends on the precise behavior of each modes in an expanding universe (recall that there are also winding modes of M2 and M5-branes in addition to the standard momentum modes). This is a delicate balancing problem and needs a more careful analysis to investigate the proper behavior of the system. In any case, even if we managed to resolve the issue, the problem of the un-cancelled vacuum energies would still plague us.

\vskip.1in

\noindent {\it Case 2: de Sitter solution as a coherent or a squeezed-coherent state}

\vskip.1in

\noindent The unsurmountable problems that we encountered by viewing \eqref{pyncmey2} as the background {\it vacuum} solution might have a simpler resolution if we view the IIB de Sitter solution, or its M-theory up-lift (as in eq (3.3) of \cite{deft}), not as a vacuum solution, but as a coherent (or a squeezed-coherent) state over a solitonic supersymmetric vacuum solution. The solitonic background solution may be taken as
in eq (2.3) of \cite{deft}, which is time-independent and therefore requires background G-fluxes to solve all the equations of motion etc \cite{BB1, DRS1, becker2}. Being supersymmetric, the vacuum energy problem is automatically resolved, and one can now quantize the modes over such a classical configuration. These modes are time-independent and their spatial behavior are typically governed by an underlying class of differential equations that take the form of Schr\"odinger equations with potentials that depend on the background solitonic data (see section (2.1) of \cite{deft} for details).  

The coherent states over such a configuration may now be easily constructed by using the creation operators for the modes as shown in \cite{deft}. Since the final outcome is required to be a solution of the form eq. (3.3) in \cite{deft} with G-fluxes as in \eqref{frostgiant2}, the coherent state construction will reproduce this as the 
most {\it probable} configuration in the Hilbert space. If we instead allow squeezed-coherent states, the probability amplitude will be even more sharply peaked over the required de Sitter configuration (although we should now take into account the dynamical behavior of the squeezed-coherent states in the Hilbert space). In any case, such a procedure reproduces the required de Sitter background as a most probable outcome in the Hilbert space. 

The first advantage of using coherent states is their close correspondence with appropriate classical counterparts. However existence of such states would still require us to solve the Schwinger-Dyson type equations, which in some sense means solving the supergravity equations of motion. If the supergravity equations do not have any solutions, then the corresponding coherent states cannot exist. The reason is that such states are constructed in an interacting theory, and therefore small fluctuations over the corresponding field configurations should directly lead to the Schwinger-Dyson equations in the path integral formalism of the theory. The good thing is, if the background \eqref{pyncmey2} (or its M-theory uplift) solve the equations of motion with the quantum corrections, then the coherent states will automatically solve the Schwinger-Dyson equations.

The second advantage of using coherent states is their ability to provide a well defined Wilsonian effective action because now the modes over which we integrate are determined from the background solitonic configuration. The latter is supersymmetric and time-independent and therefore doesn't appear to have the issues that we faced earlier regarding red-shifted modes. The modes associated to the fields that break four-dimensional de Sitter isometries may be given small masses to facilitate their integrating out procedures.

\subsection{Local quantum effects in de Sitter space and EFT  description \label{local}}

In whatever ways the high-energy modes are integrated out, we expect that at certain energy scales the quantum terms may be expressed generically as \eqref{phingsha0}.
Additionally, 
each of the quantum terms of the form \eqref{phingsha0} is suppressed by a specific power of $M_p$ that depend on the choice of ($l_i, n$) in \eqref{phingsha0}. This may be determined as:
\bg\label{kamagni2}
\sigma(\{l_i\}, n_1, n_2) = n_1 + n_2 + 2\sum_{i = 1}^{27} l_i + \sum_{k = 28}^{38}l_k, \nd
where we see that, because of the involvement of large number of terms, there could be a certain level of degeneracies.  Such degeneracies are inevitable in any quantum system, so there is no surprise that we have them here too. On the other hand, the $g_s$ scalings of the quantum terms take the form $g_s^{\theta_k}$, where: 
\bg\label{melamon0}
\theta_k & = & {2\over 3} \sum_{i = 1}^{27} l_i + {n_0 + n_1 + n_2\over 3} + {1\over 3}\left(l_{34} + l_{35}\right) + 
{2\over 3}\left(k_1 + 2\right)\left(l_{28} + l_{29} + l_{31}\right) \nonumber\\
& & -{2n_3\over 3} +  {1\over 3}\left(2 k_1 +{1}\right)\left(l_{30} + l_{32} + l_{33}\right) + 
{2\over 3}\left(k_2 -{1}\right)\left(l_{36} + l_{37} + l_{38}\right), \nd
with ($k_1, k_2$) being the two possible set of modings $k$ that appear in \eqref{frostgiant2}. The former is related to all the G-flux components other than ${\bf G}_{MNab}$ and the latter is related to the remaining three G-flux components ${\bf G}_{MNab}$. Clearly, as discussed above $k_1 \ge 0, 
k_2 \ge {3\over 2}$, and this keeps the $g_s$ scaling $\theta_k > 0$ provided $n_3 = 0$.  The quantities $n_0$ and $n_3$ are new and are related to the number of temporal derivatives and derivatives along the toroidal directions ${\mathbb{T}^2\over 
{\cal G}}$ respectively, whereas ($n_1, n_2$), that appeared in \eqref{phingsha0} earlier, are derivatives along 
${\cal M}_4$ and ${\cal M}_2$ respectively. The temporal derivatives don't change the form of 
\eqref{phingsha0} because $\partial_0 \propto {\partial\over \partial g_s}$, so $n_0$ can be absorbed by shifts in ($n_1, n_2$). Clearly positivity of $\theta_k$ is related to the background fluxes and metric being {\it independent} of the toroidal directions. 

Let us pause for a moment to grasp the significance of what we said above. The fact that the G-flux modings and the derivative constraint keep $\theta_k > 0$ is important. What it tells us is that, for any given values of $\theta_k$, there are only finite choices of ($l_i, n$) as both $l_i \in \mathbb{Z}, n_i\in \mathbb{Z}$, i.e no fractional powers of fluxes and curvature tensors are allowed here\footnote{It will be hard to assume how in our Wilsonian action fractional powers could arise.}. 
In fact this also means $\theta_k \in 
{\mathbb{Z}\over 3}$, where note the appearance of the $1/3$ factor again. Any relative {\it minus} sign will automatically imply an infinite choices for ($l_i, n$) for any given values of $\theta_k$. Such relative minus signs appear if we make $k_1 = k_2 = 0$ and $n_3 > 0$, i.e make the G-flux components time 
{\it independent}, and assume flux and curvature dependences on all the coordinates of the eight-manifold in M-theory. In other words, the $g_s$ scalings in the absence of any time-dependent G-fluxes become 
$g_s^{\theta_0}$, where:

{\footnotesize
 \bg\label{kkkbkb3}
 \theta_0  =  {2\over 3}  \sum_{i = 1}^{27} l_i + {n_0 + n_1 + n_2 \over 3} - {2n_3\over 3} + 
 {1\over 3}\left( l_{30} + \sum_{p = 1}^{4} l_{31 + p}\right)  
 + {4\over 3}\left(l_{28} + l_{29} + l_{31}\right) -{2\over 3}\sum_{q = 1}^3 l_{35 + q}, \nd} 
where even if we assume derivative constraint, switching on any components of ${\bf G}_{MNab}$ would trigger an avalanche of quantum terms\footnote{Including KK modes, but we will not discuss them here. See \cite{deft} for details on this.}. Of course since all these terms are suppressed by different powers of 
$M_p$ from \eqref{kamagni2} $-$ ignoring possible finite degeneracies $-$ the question is whether such a proliferation of quantum terms essentially spells a collapse of EFT here. Couldn't we take $M_p \to \infty$ and get rid of terms suppressed by higher powers of $M_p$, even if there are an infinite number of terms for any given choice of $\theta_0$? 
 
The answer, as elaborated in \cite{maxpaper}, is unfortunately not. 
To see this, we need the $g_s$ scalings of 
not the Lorentz invariant potential \eqref{ducksoup2}, but the quantum energy momentum tensors. In fact, what we need are the energy momentum tensors along the $2+1$ dimensional space-time, the internal six manifold ${\cal M}_4  \times {\cal M}_2$ and the toroidal directions. Their respective scalings have been worked out in \cite{deft} and we can express them as:
\bg\label{charaangul2}
\theta_k ~\rightarrow~\left(\theta_k - {8\over 3}, ~\theta_k - {2\over 3}, ~\theta_k + {4\over 3}\right), \nd
where we see that the changes are minimal: there are only constant shifts. Thus, for example, if we want to write the equations of motion along the $2+1$ dimensional space-time, we are looking for $g_s$ neutral terms (i.e the Einstein tensor to the zeroth order in $g_s$). These appear from the quantum terms in 
\eqref{phingsha0} that scale as:
\bg\label{chukkaM}
\theta_k = {8\over 3}, \nd
and therefore involve quartic order polynomials in curvatures or eight derivative polynomials in fluxes (or a mix of both with derivatives). As long as $\theta_k$ have no relative minus signs, i.e when the G-fluxes are time-dependent, the RHS of equations of motion will have finite number of quantum terms. Once we make the fluxes time-independent, the RHS of the Einstein equations will now have an infinite number of terms
going as:
\bg\label{evader} 
\sum_i {{\cal O}^{(i)}_{MN}\over M_p^{\sigma_i}} = \sum_{\{l_i\}, n}{a_1\over M_p^{\sigma(\{l_i\}, n)}}
\left(a_2 {\bf g}_{MN} \mathbb{Q}_T^{(\{l_i\}, n)} + {\partial \mathbb{Q}_T^{(\{l_i\}, n)} \over \partial {\bf g}^{MN}}\right), \nd
where ($a_1, a_2$) are constants that only depend on the dimensions of space-time; and a specific choice for $\sigma_i$ would fix a {\it finite} set of values for ($\{l_i\}, n$) from \eqref{kamagni2}. The other quantities appearing above are the warped metric ${\bf g}_{MN}$ and the quantum terms $\mathbb{Q}_T^{(\{l_i\}, n)}$
from \eqref{phingsha0}. The upshot of \eqref{evader} then is that the operators 
${\cal O}_i$ are related to the sum of all the quantum terms whose ($\{l_i\}, n$) values are fixed by $\sigma_i$ from \eqref{kamagni2}.  One may also confirm the $g_s$ scalings \eqref{charaangul2} from the 
RHS of \eqref{evader}.

As explained in \cite{maxpaper}, the infinite quantum terms are balanced against two classical set of terms: the  Einstein tensor and the energy-momentum tensors for the G-fluxes. Such a system has no hope for a solution unless ${\cal O}_i$ are of the same order as $M_p$. Unfortunately, allowing this implies our inability to ignore {\it any} term in the series, and since the series is infinite, there is no simple way we can define an EFT here! The conclusion remains unaltered even if we go to any arbitrary order in $g_s$. 

Let us elaborate this in some more details here. The key lies in the observation that, in the absence of any time-dependent (or $g_s$ dependent) G-fluxes, the relative minus signs come from two set of terms: the 
${\bf G}_{MNab}$ flux components and the number of derivatives along the ($x_3, x_{11}$) directions. The G-flux ansatze that is relevant for our case has already appeared in \cite{deft} and here we write this as:
\bg\label{canyon}
{\bf G}_{MNab} = {\bf F}_{MN}(y) \otimes \Omega_{ab}(y^a) + {1\over 3!} \partial_{a} \Big({\bf C}_{MNb}(y) \otimes
\Phi(y^a)\Big) \left(\delta_{a,11} \delta_{b3}  + \delta_{a3} \delta_{b, 11}\right), \nd
where $\Phi(y^{a, b})$ is associated with the KK modes along the toroidal directions as elaborated in 
\cite{deft}. We can ignore $\Phi(x_3)$ from outset because it will interfere with the Busher's rules that connect us to the IIB picture. The other scalar function $\Phi(x_{11})$ typically takes the form 
$\Phi(x_{11}) = {\rm cos}\left({\vert n\vert x_{11} \over {\bf R}_{11}}\right)$ with  ${\bf R}_{11}$ being the warped\footnote{By {\it warped} we will henceforth always mean $g_s$ dependent, unless mentioned otherwise.} 
eleven-dimensional radius, and $n \in \mathbb{Z}$. However since ${\bf R}_{11} \propto g_s^{2/3}$, we are effectively looking at the KK modes that go as 
${\rm cos} \left(\vert n \vert g_s^{-2/3}\right) = {\rm cos}\left(\vert n \vert \infty\right)$ near $g_s \to 0$. This doesn't have a simple perturbative expansion, in addition to being $g_s$ dependent, 
and therefore we will restrict ourselves with the $n = 0$ modes, and view the other modes to be fixed non-perturbatively. This then leaves us with only the first term, where $\Omega_{ab}(y^a)$ is a localized two-form and ${\bf F}_{MN}(y)$ is the gauge field on a charge-neutral set-up of IIB seven-branes as in 
\cite{senMM}.

We want to see how such a choice of G-flux components influence the dynamics of the system. We will start by discussing the time-independent case by making the G-flux components independent of $g_s$. To proceed, let us define the following three variables:
\bg\label{melmay}
&&\mathbb{N}_1 \equiv  l_{28} + l_{29} + l_{31}, ~~~ \mathbb{N}_2 \equiv 
l_{36} + l_{37} + l_{38} \nonumber\\
&&\mathbb{N}_3 \equiv {2} \sum_{i = 1}^{27} l_i + {n_0 + n_1 + n_2} + l_{34} + l_{35} + l_{30} + l_{32} + l_{33}, \nd
where $(\{l_i\}, n_j) \in (\mathbb{Z}, \mathbb{Z})$ are as defined in \eqref{phingsha0}. The above definitions
are useful to study the time-independent cases, but they do 
rely on what lower bounds are placed on ($k_1, k_2$) in the time-dependent cases. If $k_i \ge {3\over 2}$, as taken in \cite{deft}, then 
$\mathbb{N}_3$ may be further subdivided into $\mathbb{N}_3^{(a)}$, containing the last three terms, and 
$\mathbb{N}_3^{(b)}$, containing the remaining five terms. For the present discussions, we will avoid this subdivision so that $k_1 = 0$ and $k_2 = {3\over 2}$ serve as the lower bounds for the time-dependent cases and $k_1 = k_2 = 0$, for the time-independent cases. For the latter, if we want to know how many operators are possible to order ${g_s^{a_1/3}\over M_p^{b_1}}$, the answer lies in the following set of equations:
\bg\label{wynn}
\mathbb{N}_1 + \mathbb{N}_2 + \mathbb{N}_3 + n_3 = b_1, ~~~
4\mathbb{N}_1 -2 \mathbb{N}_2 + \mathbb{N}_3 -2 n_3 = 8 + a_1, \nd
where for simplicity we have restricted ourselves to the $2+1$ dimensional space-time EOMs. The quantity
$n_3$ denotes the number of derivatives along the $(a, b) = (x_3, x_{11})$ directions, and to respect the Buscher's rule, we will only consider derivatives along the eleventh-direction. The first equation in 
\eqref{wynn} is related to the $M_p$ scalings of the quantum terms \eqref{phingsha0}, and we start by assuming that ${\bf G}_{MNab}$ do not have any inherent $M_p$ dependences. The second equation now measures the $g_s$ scalings and is derived from \eqref{kkkbkb3}. The relative minus sign is important here. It tells us that, for a given choice of $a_1$, there are in fact an infinite choices of $\mathbb{N}_i$ and 
$n_3$. We can try to simplify the above set of equations as:
\bg\label{riobuffet}
\mathbb{N}_1 - \mathbb{N}_2 - n_3 = {8 + a_1 - b_1\over 3}, ~~~~ 2\mathbb{N}_1 + \mathbb{N}_3 = 
{8 + a_1 + 2b_1\over 3}, \nd
but the conclusion remains unchanged: there cannot be finite number of local quantum terms in the case when the fluxes are time-independent. All these terms are suppressed by various powers of $M_p$, so the question is whether we can allow for $M_p$ hierarchy by simply making $M_p \to \infty$. Note that we cannot make $b_1 = 0$ in \eqref{wynn} without trivializing everything, although $a_1 = 0$ is always possible. Thus the standard analysis allows $g_s$ neutral terms, but not $M_p$ neutral terms. In other words 
the $g_s^0$ parts of the LHS of Einstein's equations are suppressed by zeroth power of $M_p$ (and so are the classical contributions to these equations), but the RHS, that have the quantum terms, are suppressed by $M_p^{b_1}$. This may seem like a way out: making $M_p \to \infty$ we can get rid of most of the infinite quantum terms. 

However the situation is more involved because we have ignored all inherent $M_p$ dependences from the localized G-fluxes itself. Where would the $M_p$ dependences come from? The answer lies in our choice of the localized two-form $\Omega_{ab}$, for which we make the following ansatze:
\bg\label{prestige}
\Omega_{ab}(x_{11}) = \sum_{n = 1}^\infty {\cal A}_n {\rm exp}\Big(-M_p^{2n} x^{2n}_{11}\Big) \epsilon_{ab}, \nd
where ${\cal A}_n$ are the normalization constants
that {\it do not} depend on $M_p$. In the above series, the exponential series dies off faster as $n$ increases, so we can suffice ourselves by choosing only the smallest, i.e $n = 1$, term. Plugging \eqref{prestige}, with $n = 1$, in \eqref{canyon} and then in 
\eqref{phingsha0} will change the first equation in \eqref{wynn} to the following:
\bg\label{fatliv}
\mathbb{N}_1 + \mathbb{N}_2 + \mathbb{N}_3 - n_3 = b_1, \nd
where the relative minus sign comes from the fact that every $n_3$ derivatives will now contribute
$M_p^{-2n_3}$ to the quantum series \eqref{phingsha0}. We can now make $a_1 = b_1 = 0$ in \eqref{fatliv} and in the second equation of \eqref{wynn}. This gives us the condition:
\bg\label{rusmey}
2\mathbb{N}_1 - 4\mathbb{N}_2 - \mathbb{N}_3 = 8, \nd
that allows for an infinite number of solutions at zeroth orders in $g_s$ and $M_p$. These quantum terms then contribute to the RHS of the Einstein's equations spoiling the EFT description. 

The above representation of the localized two-form \eqref{prestige} isn't the only available choice. 
In fact one might question the spread of the gaussian \eqref{prestige} itself when we shrink the torus to zero size to go to the IIB limit. To avoid such conundrums, let us consider another choice of the two-form given by:
\bg\label{misangry} 
\Omega_{ab}(y^\alpha) = \sum_{n = 1}^\infty {\cal A}_n {\rm exp}\Big[-\left(g_{\alpha\beta} 
y^\alpha y^\beta\right)^n M_p^{2n}\Big] \epsilon_{ab}, \nd
where $g_{\alpha\beta}$ is the {\it un-warped} metric that appears in the M-theory lift of \eqref{pyncmey1} or \eqref{pyncmey2}. This way no $g_s$ dependences are introduced in the definition of the G-flux components. The choice \eqref{misangry} tells us that the four-manifold, with local Taub-NUT structure, is 
${\cal M}_2 \times {\mathbb{T}^2\over {\cal G}}$ and we can use coordinates $y^\alpha \equiv (y^4, y^5)$ to  
represent ${\cal M}_2$. This way the four-manifold has coordinates ($x_3, x_{11}, y^4, y^5$) and the type IIB seven-brane will span the four-dimensional space-time (with de Sitter isometries) and wrap the {\it other} four-manifold ${\cal M}_4$. 

We can also ask how \eqref{rusmey} changes under the choice of the two-form \eqref{misangry}. To proceed, let us define $\mathbb{N}_3$ in \eqref{melmay} as $\mathbb{N}_3 \equiv \mathbb{N}_4 + n_2$, with $n_2$ being the number of derivatives along ${\cal M}_2$ in \eqref{phingsha0}. The derivative actions now pull out 
$M_p^{-2n_2}$, thus changing \eqref{rusmey} to:
\bg\label{ramrav}
5\mathbb{N}_1 - \mathbb{N}_2 + 2\mathbb{N}_4 - n_3 = 8, \nd
implying, as before, an infinite number of possible terms to zeroth orders in $g_s$ and $M_p$. The conclusion then remains the same: there is again no EFT description possible with time-independent G-flux components. 

Switching on time dependences miraculously cures the problem as may be see from the following example.
One of the simplest change we can make in our G-flux ansatze \eqref{canyon}, is to insert {\it warped} metric in the definition of the localized two-forms \eqref{prestige} and \eqref{misangry}. This will convert these two-forms to:
\bg\label{alexmey}
\Omega_{ab}(y^c, y^\alpha) &=& \sum_{n, k = 0}^\infty {\cal A}_{nk}~ {\rm exp}\Big[-\left({\bf g}_{cd} 
y^c y^d\right)^n M_p^{2n}\Big] {\rm exp}\Big[ -\left({\bf g}_{\alpha\beta} y^\alpha y^\beta\right)^k M_p^{2k}\Big] \epsilon_{ab}\\
& = & \sum_{n, k = 0}^\infty {\cal A}_{nk}~ {\rm exp}\Big[-\left(g_s^{4/3} {g}_{cd} 
y^c y^d\right)^nM_p^{2n}\Big] {\rm exp}\left[ -\left({{g}_{\alpha\beta} y^\alpha y^\beta \over g_s^{2/3}}\right)^k M_p^{2k}\right] \epsilon_{ab}, \nonumber \nd
where we take ${\cal A}_{00} = 0$, and ${\cal A}_{nk}$ is only function of $g_s$ but not of $M_p$. The above representation of the two-form \eqref{alexmey}, when plugged-in in the first term of \eqref{canyon}, provides the G-flux components that resemble closely our original ansatze \eqref{frostgiant2} for the time-dependent flux components. In the late-time limit when $g_s \to 0$, the second term that goes as 
${\rm exp}\left(-{1\over g_s^{2/3}}\right)$ approaches zero faster
than any polynomial powers of $g_s$. This decouples for $k = 0$, and therefore our localized two-form becomes:
\bg\label{plazabingo}
 \Omega_{ab}(y^c) = \sum_{n = 1}^\infty {\cal A}_{n0}~ {\rm exp}\Big[-\left(g_s^{4/3}{\delta}_{cd} 
y^c y^d\right)^nM_p^{2n}\Big] \epsilon_{ab}, \nd
and as before we will concentrate only on $n = 1$ as other polynomial powers die off faster. 
However before moving further, we want to discuss the form of the normalization constant ${\cal A}_{nk}$ that we kept under the rug so far. First, do we really need to normalize \eqref{alexmey} or \eqref{plazabingo}? Couldn't we keep ${\cal A}_{nk}$ as an arbitrary function of $g_s$ but independent of $M_p$? To analyze this, let us first assume that normalization is necessary. We take the most general two-form, i.e 
\eqref{alexmey}, and normalize this according to the following prescription:
\bg\label{millenium}
M_p^4 \int_{-\infty}^{+\infty} d^4 z \sqrt{{\bf g}_{(4)}}~ {\bf g}^{ac} {\bf g}^{bd}~ \Omega_{ab} \Omega_{cd} = 1, \nd
where $z$ denotes the coordinates of the four-manifold ${\mathbb{T}^2\over {\cal G}} \times {\cal M}_2$, i.e $z \equiv (x_3, 
x_{11}, y^\alpha, y^\beta)$, ${\bf g}_{ab}$ is the warped metric of ${\mathbb{T}^2\over {\cal G}}$, and 
${\bf g}_{(4)}$ is the determinant of the metric of the four-manifold. Note that we have used the {\it non-compact} limit of the four-manifold to apply our normalization scheme.
The two-form $\Omega_{ab}$ in 
\eqref{alexmey} typically involves a sum over an infinite number of terms in ($n, k$), but one can show that 
in general ${\cal A}_{nk} = c_{nk} g_s^{4/3}$, where $c_{nk}$ are constants independent of $M_p$ and 
$g_s$, and only depend on the choice of ($n, k$). In deriving this, we have also assumed that the un-warped metric components $g_{ab}$ and $g_{\alpha\beta}$ may be approximated by $\delta_{ab}$ and 
$\delta_{\alpha\beta}$ respectively. Of course in our assumption $g_{ab} = \delta_{ab}$ always, so if $g_{\alpha\beta}$ is not exactly $\delta_{\alpha\beta}$, the analysis will get slightly more complicated, but in the end, any changes to the $g_s$ scalings of ${\cal A}_{nk}$ wouldn't change the final outcome as we shall see below. Interestingly, with this $g_s$ scalings of ${\cal A}_{nk}$, the G-flux components, by plugging in \eqref{alexmey} in \eqref{canyon}, matches precisely with the G-flux ansatze that we took in \eqref{frostgiant2} with $n = 0, k = 2$ (($n, k$) in \eqref{frostgiant2} not to be confused with 
($n, k$) in \eqref{alexmey}). Clearly, as emphasized in \cite{deft}, such a choice of G-flux components will lead to a four-dimensional EFT description. 

What happens in the $g_s \to 0$ limit? As mentioned above the two-form can be simplified to 
\eqref{plazabingo} at all points except at the origin where $y^\alpha y_\alpha = 0$. In this case the second term in \eqref{alexmey} becomes 1, so \eqref{plazabingo} with coefficients ${\cal A}_{n0}$ does capture the simplified case.  The normalization condition, in the non-compact limit,  now becomes: 
\bg\label{salander}
M_p^2 \int_{-\infty}^{+\infty} d{x_3} dx_{11}  \sqrt{{\bf g}_{(2)}}~ {\bf g}^{ac} {\bf g}^{bd}~ \Omega_{ab} \Omega_{cd} = 1, \nd
where ${\bf g}_{(2)}$ is the determinant of the toroidal metric of ${\mathbb{T}^2\over {\cal G}}$.
This way again
${\cal A}_{n0} = c_{n0} g_s^{4/3}$, and as mentioned earlier we can take $n = 1$ in \eqref{plazabingo} so that ${\cal A}_{nk} = 0$ for $n > 1, k \ge 0$. Note that again the G-flux components satisfy the condition laid out in \eqref{frostgiant2} and therefore according to \cite{deft} we expect an EFT to exist. 

Let us verify the above statement more explicitly. For definiteness, we will take the two-form 
\eqref{plazabingo} with $n = 1$, so that the late time $g_s \to 0$ limit remains finite. 
Now once we make
$n_3$ derivatives, we get an extra factor of $g_s^{4n_3/3} M_p^{2n_3}$, implying that the $g_s$ scaling changes as $g_s^{4n_3/3}$, whereas the $M_p$ scaling changes as 
$M_p^{-2n_3}$. Plugging-in these changes to \eqref{wynn} would convert \eqref{rusmey} and \eqref{ramrav}
to:
\bg\label{paulamey}
6\mathbb{N}_1 + 4\mathbb{N}_2 +  3\mathbb{N}_3 = 8, \nd
where the normalization constant ${\cal A}_{10}$ contributes the extra ${4\mathbb{N}_2\over 3}$ factor.
The equation \eqref{paulamey} now has only a finite number of solutions, thus allowing an EFT description to exist at zeroth orders in $g_s$ and $M_p$. 
In fact \eqref{paulamey} is {\it always} true 
for any choice of $n \ge 1$ (and $ k = 0$) in \eqref{plazabingo}, implying that EFT description exists once time-dependences are switched on. One could also introduce derivative constraint by making all fields independent of the toroidal directions, as in \cite{deft}, and then use the time-dependent G-flux ansatze 
\eqref{frostgiant2}. The results remains the same: EFT description exists for the time-dependent fluxes, but 
immediately goes away once the time-dependences are switched off.

The result \eqref{paulamey} also gives us a hint why normalization may be necessary. In the absence of any normalization scheme, there would be no $4\mathbb{N}_2$ piece in \eqref{paulamey}, and therefore arbitrary powers of ${\bf G}_{MNab}$ may be switched on at zeroth order in $g_s$ and $M_p$, ruining an EFT description. The coefficient of $\mathbb{N}_2$ is solely determined by the $g_s$ scaling of 
${\cal A}_{nk}$, and therefore as long as we have a positive scaling satisfying the bound in \eqref{frostgiant2}, EFT description would {\it always} be possible even if $g_{\alpha\beta}$ is not exactly 
$\delta_{\alpha\beta}$.
One may note that, both the choices of the two-forms in \eqref{alexmey} and 
\eqref{plazabingo} with warped metric and the normalization condition are a {\it natural} way to attain the G-flux ansatze of \eqref{frostgiant2}. Without normalization or without time-dependences, ${\cal A}_{nk}$ 
in \eqref{alexmey} would remain $g_s$ independent, and thus creating problems with EFT descriptions.

The above conclusion for time-independent fluxes, despite its negative connotation, is an expected consequence for any theory that allows an infinite set of operators at any given  order in $g_s$. This however does not quite preclude a UV completion of the model, and therefore we will call such breakdown of an EFT as the {\it standard breakdown}. 
Unfortunately the story does not end here, and the system has an even more severe breakdown of EFT once time-dependences are switched-off. We will call the second kind of breakdown as the {\it swampland breakdown}, and has its root in the non-local quantum terms. This is the discussion that we turn to next. 


\section{Analysis with non-local quantum terms \label{qotom2}}

The non-local quantum terms are known to exist in M-theory and some aspects of it can be seen directly in IIA using string field theory \cite{calagni2}. In fact the simplest form of non-local interactions already appear from \eqref{phingsha0} in the limit $n_i \to \infty$, where 
$n_i$ are the number of derivatives along specific directions on the internal eight-manifold as mentioned earlier. 
Another set of non-local interactions may be expressed again in terms of the local quantum terms \eqref{phingsha0} in the following suggestive way: 
\bg\label{koliman2}
\mathbb{W}^{(\{l_i\}, n)} \equiv
\left(\sum_{q = 1}^{\infty} {C_q M_p^{2q}\over \square^q}\right)\mathbb{Q}_{\rm T}^{(\{l_i\}, n)}, \nd
where $C_q$ in principle be functions of all the coordinates of the internal eight-manifold, where we have defined the $\square$ operator, as well as of $g_s$. The above set of non-local interactions are seen from the local interactions \eqref{phingsha0} by making $(n_1, n_2, n_3) \to (-n_1, -n_2, -n_3)$. The $g_s$ scaling then typically picks up a factor of the form $g_s^{\theta'_k}$ with:
\bg\label{suspiria}
\theta'_k = {2n_3 - n_0 - n_1 - n_2\over 3}, \nd 
where taking $n_0 = 0, n_1 = 4r, n_2 = 2r$ and $n_3 = 2r$, we get $\theta'_k = -{2r\over 3}$ which is exactly the extra $g_s$ scaling appeared in section 3.2.6 of \cite{deft} if we identify $r$ to the level of non-locality (we will take $r \ge 1$ so as to avoid taking $r = 0$ case). For $n_0 \to -n_0$ we get temporal non-locality.
This doesn't change the story (see \cite{deft}) so we will avoid discussing this.

Introducing non-local interactions however is not that straight-forward because we do expect these non-localities to play no part in the time-independent supergravity limit. At high energies, again in the time-independent case, the non-local interactions should somehow be manifested one way in {\it fattening} up the lump solutions from string field theory. Lump solutions from string field theory have been discussed for some simple cases in \cite{raduL}, and we expect non-localities to show up in their sizes. More complicated form of non-localities are also possible, but these have been mostly unexplored as the high energy limit of M-theory is still an unknown territory. In fact it is possible to construct an example of a non-local interaction that completely shuts off in the standard supergravity limit.

In the time-dependent cases we don't know the generic structure of the non-local interactions, but we do expect that a class of them may be represented by inverse derivative actions of the form \eqref{koliman2}.
We can even convert the inverse derivatives to integrals, and re-express \eqref{koliman2} in the following way:  
\bg\label{pmoran2}
\mathbb{W}^{(1)}(y) \equiv \mathbb{W}^{(\{l_i\}, n; 1)} 
= \int d^8 y' \sqrt{{\bf g}_8} \left({ \mathbb{F}^{(1)}(y - y') 
\mathbb{Q}_{\rm T}^{(\{l_i\}, n)} (y') \over M_p^{\sigma(\{l_i\}, n) - 8}}\right),  
 \nd
where note that we have used a non-locality function $\mathbb{F}^{(r)}(y - y')$ with $r = 1$ being the lowest order of non-locality. These non-locality functions should become sharply peaked functions in the standard supergravity limits, so that eleven-dimensional supergravity may be defined in the standard way with local interactions. The general $r$-th order non-local interactions may be constructed iteratively as:
\bg\label{karalura2} 
\mathbb{W}^{(r)}(y) &=& M_p^8 \int d^8y' \sqrt{{\bf g}_8(y')} ~\mathbb{F}^{(r)}(y - y') \mathbb{W}^{(r-1)}(y')\\
& = & M_p^{16} \int d^8 y' \sqrt{{\bf g}_8(y')}~\mathbb{F}^{(r)}(y - y') 
\int d^8y'' \sqrt{{\bf g}_8(y'')} ~\mathbb{F}^{(r-1)}(y' - y'') \mathbb{W}^{(r-2)}(y''). \nonumber
 \nd  
The above way of expressing the non-local interactions show that not only the $M_p$ scalings change to every order in $r$, but also the $g_s$ scalings, because the non-locality function $\mathbb{F}^{(r)}(y - y')$ could in principle have explicit dependence on $M_p$ and $g_s$ also. Let us then assume that the most 
{\it dominant} contribution to the non-locality function $\mathbb{F}^{(r)}(y - y')$ may be written as:

{\footnotesize
\bg\label{spivibhalo2}
\mathbb{F}^{(r)}(y - y') = \left(M_p^6 \mathbb{V}_6\right)^{\sigma_1(r) - \sigma_1(r - 1) - 1} \left(M_p^2 
\mathbb{V}_{\mathbb{T}^2}\right)^{\sigma_2(r) - \sigma_2(r - 1) - 1} 
\left({g_s\over H}\right)^{\sigma_3(r) -\sigma_3(r - 1) + {2\over 3}} f(y - y'), \nd}
where $\mathbb{V}_6$ is the un-warped volume of ${\cal M}_4 \times {\cal M}_2$, 
$\mathbb{V}_{\mathbb{T}^2}$ is the un-warped volume of the toroidal fibre; $\sigma_i(r) \in \mathbb{Z}$ at any order in non-locality and $\sigma_i(0) = 0$ for $r \ge 1$. Finally we take $f(y-y')$ to be independent of $M_p$ and $g_s$.

\subsection{Various limits of non-localities and EFT description}

The above way of isolating the $M_p$ and the $g_s$ dependence is generic and doesn't quite depend on what specific form the non-locality function takes. The factors of $2/3$ and 1 are there to simply remove other superfluous $g_s$ and volume dependences so that all $M_p$ and $g_s$ dependences are accounted for by ($\sigma_1(r), \sigma_2(r)$) and $\sigma_3(r)$ respectively. This repackaging is just for the sake of convenience and therefore bears no consequence on the final outcome. In the following, let us study the behavior of the non-local quantum interactions for various limits of $g_s$ and $M_p$.

\vskip.1in

\noindent {\it Case 1: The $g_s$ scalings do not change, $M_p$ scalings change}

\vskip.1in

\noindent Once we plug in the dominant contribution \eqref{spivibhalo2} in \eqref{karalura2}, we expect the quantum terms to have both $M_p$ and $g_s$ dependences in addition to the $M_p$ and $g_s$ dependences from the local quantum terms \eqref{phingsha0}. Therefore when we say $g_s$ scalings do not change, we simply mean that the $g_s$ scalings fully originate from the $g_s$ scalings of the local quantum terms, i.e $g_s^{\theta_k}$ with $\theta_k$ as in \eqref{melamon0}. This means we can consider:
\bg\label{bjosie}
\sigma_2(r) =  \sigma_2(r - 1), ~~~ \sigma_3(r) =  \sigma_3(r - 1). \nd
The first condition removes any extra dependences on toroidal volume $\mathbb{V}_{\mathbb{T}^2}$ such that we will not be required to make $\mathbb{V}_{\mathbb{T}^2} \to 0$ to decouple the non-local terms. The second condition would make the $g_s$ scalings as $g_s^{\theta_k}$. However note that $\sigma_3$ can have two possible signs in \eqref{spivibhalo2}. The positive sign will make $\theta_k$ {\it negative} 
if $\vert\sigma_3(r)\vert$ is bigger than ${8\over 3}$. This will lead to no solutions in the time-dependent cases, so for the time being we will assume the relative minus in the exponent of $g_s$ in 
\eqref{spivibhalo2}.

\begin{figure}[h]
\centering
\begin{tabular}{c}
\includegraphics[width=0.8\textwidth]{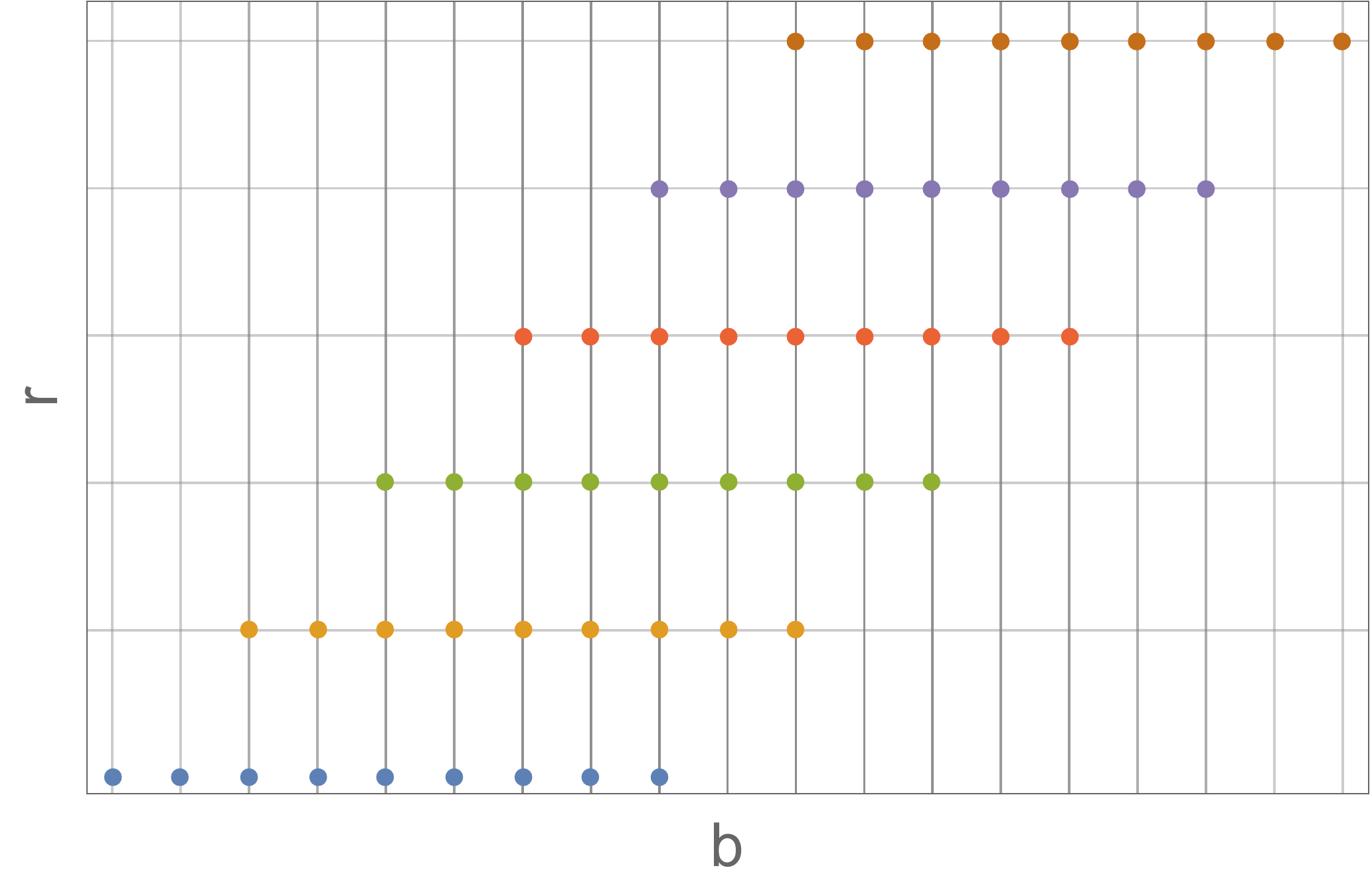}
\end{tabular}
\caption{A representation of all operators with the same $g_s$ scaling, organized according to their level of non-locality, $r$ (vertical axis) and $M_p$ scaling, $b$ (horizontal axis). In the time dependent case, each level of non-locality only has finitely many operators, each with different $M_p$ dependence. For $\sigma_3(r)=0$ changing non-locality level shifts the $M_p$ scalings, resulting in a finite number of operators on each vertical line, i.e. with a given $M_p$ scaling.}
\label{fig1}
\end{figure}

To proceed we will first take $n_3 = 0$, so the derivatives along the toroidal directions do not ruin the EFT. 
We will also take $k_1 \ge 0$ and $k_2 \ge {3\over 2}$ in \eqref{melamon0} so that $\theta_k > 0$. This way, for the time-dependent case, at least the local quantum terms provide an EFT description in the IIB side as we saw in section \ref{qotom1}. To see what happens in the presence of the non-local quantum terms, we will use the variables as defined in \eqref{melmay} before, and 
ask the following question: how many operators are allowed for a given choice of 
${g_s^{a_1/3} \over M_p^{b_1}}$ with $(a_1, b_1) \in (\mathbb{Z}, \mathbb{Z})$? We will also assume 
$a_1  \ge 0$, and hence positive, but $b_1 $ can have either signs. Lets start by taking positive $b_1 $, so that we are looking at only the quantum terms suppressed by powers of $M_p$, and from \eqref{melamon0} it is easy to see that it is the combination $\mathbb{N}_2 + \mathbb{N}_3$ that always occur together for the case with time-dependent G-fluxes. Thus we rename $\mathbb{N}_1 \equiv \mathbb{X}$ and $\mathbb{N}_2 + \mathbb{N}_3 \equiv 
\mathbb{Y}$ for convenience. We will also explore the $2+1$ dimensional space-time EOMs as they lead to the most interesting set of equations because of \eqref{chukkaM}. Combining these together, the number of quantum terms to order ${g_s^{a_1 /3} \over M_p^{b_1}}$ is governed by the following set of linear equations:
\bg\label{jukam}
4\mathbb{X} + \mathbb{Y} = 8 + a_1, ~~~~~~
\mathbb{X} + \mathbb{Y} = 6 \sigma_1(r) + b_1, \nd
where $r$ denotes the level of non-locality as mentioned earlier and $\sigma_1(r) \in \mathbb{Z}$. Clearly, by definition $(\mathbb{X}, \mathbb{Y}) \in (\mathbb{Z}, \mathbb{Z})$, and from the first equation in 
\eqref{jukam} we see that there are finite number of solutions for ($\mathbb{N}_1, \mathbb{N}_2, 
\mathbb{N}_3$). Since $g_s$ scalings do not change, there are the {\it same} number of solutions to any order in the non-locality $r$, although their $M_p$ scalings change due to the second equation in 
\eqref{jukam}. We can easily solve both the equations in \eqref{jukam} simultaneously to get:
\bg\label{shordi}
\mathbb{X} = {8 + a_1  - 6\vert \sigma_1(r)\vert - b_1 \over 3}, ~~~~~~ 
\mathbb{Y} = {24\vert\sigma_1(r)\vert + 4b_1  - 8 - a_1  \over 3}. \nd
Let us now ask what do the above equations imply. For a given choice of the non-locality $r$, the first solution provides us with the possible {\it integer} number of ($l_{28}, l_{29}, l_{31}$) that would give rise to the quantum terms scaling as ${g_s^{a_1/3} \over M_p^{b_1}}$. Similarly, the second solution provides us with possible number of integer solutions for the remaining set of ($\{l_i\}, n_j$) in \eqref{melmay}, again to the same order in ${g_s^{a_1/3} \over M_p^{b_1}}$. It is easy to see that there can only be {\it finite} number of solutions, because neither $\mathbb{X}$ nor $\mathbb{Y}$ can be negative. Since we expect $\sigma_1(r)$ to be a monotonically {\it increasing} function\footnote{This is easy to justify. To any order in $r$, 
\eqref{koliman2} and \eqref{karalura2} tell us that once we fix the lowest order (i.e $r = 1$) non-locality function $\mathbb{F}^{(1)}(y - y')$, higher order (i.e $r > 1$) non-localities are basically products of the lowest order non-locality function. This is the simplest case, but at least this suggests why $\sigma_1(r)$ can increase monotonically. If $\sigma_1(r)$ decreases monotonically, then \eqref{pmoran2} cannot be 
justified. \label{dori}}  of $r$, the positivity arguments restrict it to lie between:
\bg\label{kashi}
{8 + a_1  - 4b_1  \over 24} ~ < ~ \vert\sigma_1(r)\vert ~ < ~ {8 + a_1  - b_1  \over 6}, \nd
which immediately implies that we cannot go beyond some order in $r$ for a fixed choice of ($a_1, b_1$). Thus there can only be a finite number of operators to any order in ${g_s^{a_1/3} \over M_p^{b_1}}$. This is illustrated in {\bf figure \ref{fig1}}.

What happens when $\sigma_1(r)$ is {\it negative}? Negative $\sigma_1(r)$ implies positive power of
$M_p$, which can also naturally occur once non-localities are switched on. Such positive powers appeared 
in \cite{nodS} and were in-part responsible for the breakdown of $M_p$ hierarchies in the time-neutral series. Here introducing negative $\sigma_1(r)$ would convert $\mathbb{X}$ and $\mathbb{Y}$ from 
\eqref{shordi} to:
\bg\label{jor}
\mathbb{X}  = {8 + a_1  + 6\vert\sigma_1(r)\vert - b_1 \over 3}, ~~~~~~ \mathbb{Y} = {4b_1  - 24\vert \sigma_1(r)\vert  - 8 - a_1  \over 3}. \nd
Again, both $\mathbb{X}$ and $\mathbb{Y}$ needs to be positive, and we need the upper bound on 
$\vert\sigma_1(r)\vert$. This will now appear from the second equation in \eqref{jor}, giving us the following range for $\vert\sigma_1(r)\vert$:
\bg\label{polyps}
{b_1  - 8 - a_1 \over 6} ~ < ~ \vert\sigma_1(r)\vert ~ < ~ {4b_1  - 8 - a_1\over 24}. \nd
For fixed ($a_1, b_1$), the upper bounds clearly tells us that we cannot go to arbitrarily high level of non-localities. 
In fact the lowest order EOMs are written for zeroth order in $g_s$, i.e $a_1  = 0$, and clearly we require large $b_1$ to explore large values of $r$. Somewhat surprisingly, for the case \eqref{kashi}, the opposite in true: large $b$ makes $\mathbb{X}$ negative in \eqref{shordi}, or alternatively disallowing large $r$ in 
\eqref{kashi}. In either cases then, the number of operators are always {\it finite} to any order in 
${g_s^{a_1/3}\over M_p^{b_1}}$. See {\bf figure \ref{fig1}}.

What happens now for the case when the fluxes are completely time-independent, i.e $k_1 = k_2 = 0$ 
in \eqref{phingsha0}? To analyze this case, the ($\mathbb{X}, \mathbb{Y}$) variables are not useful, and we will have to resort back to the original ($\mathbb{N}_1, \mathbb{N}_2, \mathbb{N}_3$) variables. As before, once we fix ($a_1, b_1$), we get the following two equations:
\bg\label{jhograO}
&& 4 \mathbb{N}_1 - 2 \mathbb{N}_2 + \mathbb{N}_3 ~ = ~ 8 + a_1  \nonumber\\
&&\mathbb{N}_1 + \mathbb{N}_2 + \mathbb{N}_3 ~ = ~ 6\sigma_1(r) + b_1, \nd
where note the relative {\it minus} sign in the first equation. This already tells us that, to any order in 
$g_s^{a_1}$, there is an {\it infinite} number of integer values for ($\mathbb{N}_1, \mathbb{N}_2, 
\mathbb{N}_3$), leading to an infinite possible set of operators from \eqref{phingsha0} for $r = 0$. Since $g_s$ scalings do not change, we also expect an infinite set of operators for all $r \ge 1$. All these operators have different $M_p$ scalings for any $r$ as evident from the second equation in \eqref{jhograO}, so the question is how many operators do we expect to any order in ${g_s^{a_1/3} \over M_p^{b_1}}$?  The answer lies in solving both the equations in \eqref{jhograO} simultaneously. There are three unknowns, but only two equations, so we won't be able to solve for all them. In any case it suffices to  note that:
\bg\label{vampirella}
\mathbb{N}_1  - \mathbb{N}_2  ~ = ~ {8 + a_1  - 6\sigma_1(r) - b_1  \over 3}, \nd
where now the sign of $\sigma_1(r)$ really doesn't matter. For example if $\sigma_1(r) = 
+ \vert\sigma_1(r)\vert$, there is no upper or lower bounds on it and therefore we can go to any arbitrary level of non-localities and still get operators to order ${g_s^{a_1/3} \over M_p^{b_1}}$. When $\sigma_1(r) = 
-\vert\sigma_1(r)\vert$, the relative minus signs on both sides of the equation \eqref{vampirella} again 
guarantees an infinite number of operators to any order in ${g_s^{a_1/3} \over M_p^{b_1}}$. Switching on 
$n_3$ as in \eqref{wynn} and using the two-form \eqref{alexmey} with {\it unwarped} metric $-$ so that we are still in the time-independent scenario $-$ will allow us to go to the zeroth order in $g_s$ and $M_p$ giving:
\bg\label{amberH}
2\mathbb{N}_1 - 4\mathbb{N}_2 - \mathbb{N}_3 + 12\sigma_1(r) = 8. \nd
Because of the relative minus signs, \eqref{amberH} has an infinite number of integer solutions. 
This is therefore a clear sign that the EFT is breaking down in the time-independent case.  

\vskip.1in

\noindent {\it Case 2: The $g_s$ and the $M_p$ scalings both change}

\vskip.1in

\noindent Let us now discuss the more interesting case where both $g_s$ and the $M_p$ scalings change. 
Of course we expect that whenever $g_s$ scalings change $-$ as we go to the higher levels of non-localities $-$ the $M_p$ scalings would {\it automatically} change. Here we want to explore the additional changes to the $M_p$ scalings from non-local interactions, and therefore we take:
 \bg\label{bjosie2}
\sigma_2(r) =  \sigma_2(r - 1), \nd
to decouple the effects of the toroidal volume so that they do not effect the results. The trivial case of 
decoupling the non-localities by taking the limit of the vanishing toroidal volume was already considered in \cite{deft}. Here we want to explore the more non-trivial cases.

\begin{figure}[h]
\centering
\begin{tabular}{c}
\includegraphics[width=0.8\textwidth]{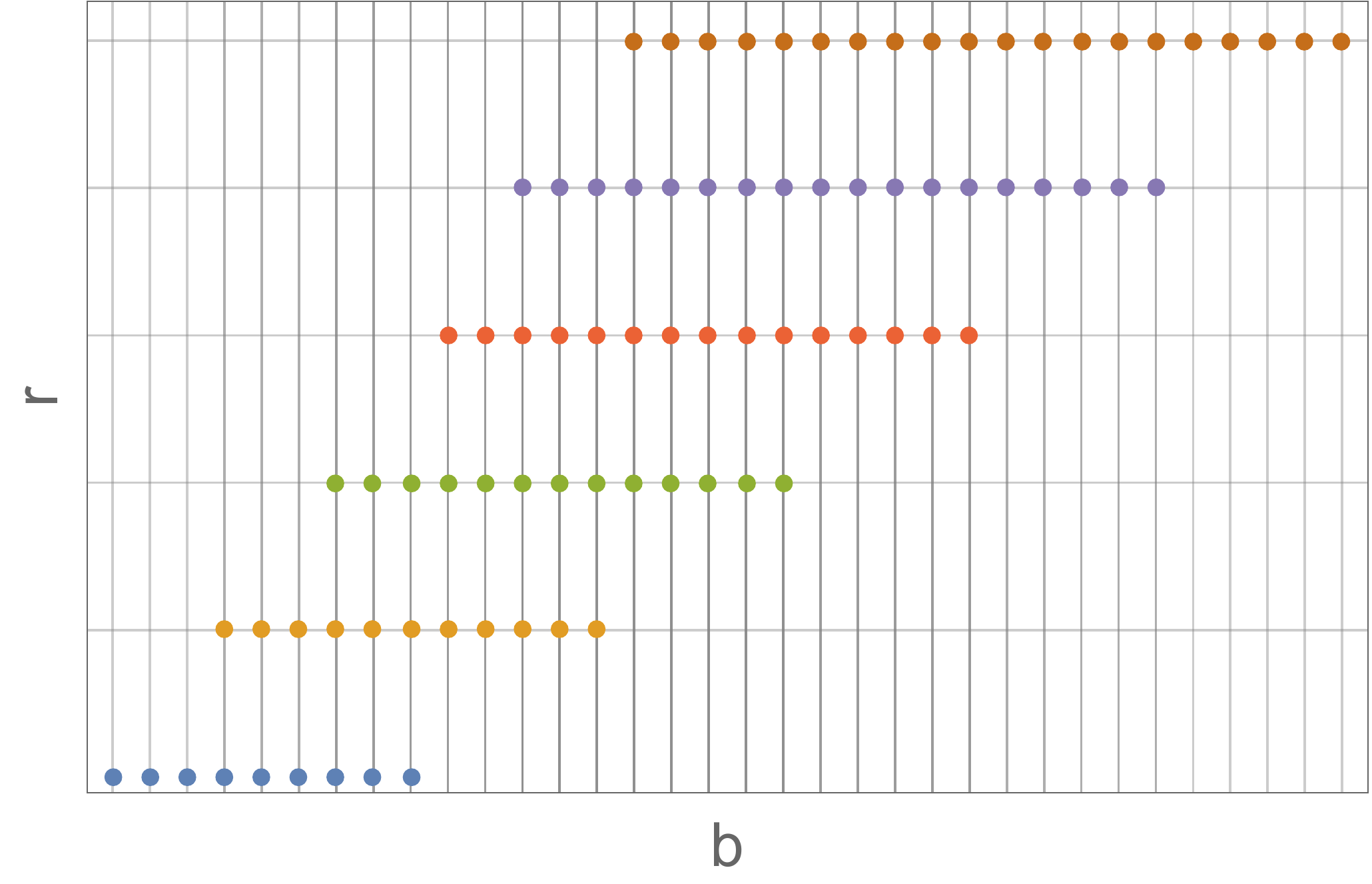}
\end{tabular}
\caption[]{For $\sigma_3(r) \neq {\rm constant}$, the number of operators with the same $g_s$ scaling can be different for different levels of non-locality. With $\sigma_1(r) > 0$ but falling outside of the window \eqref{rebhall} the number of operators at higher non-locality level proliferate faster, but there are still finitely many for fixed $M_p$ scaling.}
\label{fig2}
\end{figure}

Our starting point would be to use the $\mathbb{X}$ and $\mathbb{Y}$ variables that we used earlier, but now our equations differ from \eqref{jukam} by the introduction of $\sigma_3(r)$ associated with $g_s$ scalings. The equations become:
\bg\label{sbbooking}
4\mathbb{X} + \mathbb{Y} ~ =  -3 \sigma_3(r) + 8 + a_1, ~~~~~~ \mathbb{X} + \mathbb{Y} ~ = ~ 6\sigma_1(r) 
+ b_1, \nd
where $\vert\sigma_3(r)\vert \in \mathbb{Z}$ and we will again assume this to be a monotonically increasing function of $r$. The reasoning remains somewhat similar to what we had in footnote \ref{dori} for the case 
$\sigma_1(r)$. Of course the simplest case would be when $\vert\sigma_3(r)\vert = {2r\over 3}$ but could envision more complicated form for $\sigma_3(r)$. Interestingly, when $\sigma_3(r) = +\vert\sigma_3(r)\vert$,
the RHS of the first equation in \eqref{sbbooking} can become negative unless $\vert\sigma_3(r)\vert < 
{8\over 3}$, for $a_1  = 0$. To avoid this, we take $\sigma_3(r)  = -\vert\sigma_3(r)\vert$ in \eqref{sbbooking}, and get the following values for $\mathbb{X}$ and $\mathbb{Y}$:
\bg\label{tesla}
\mathbb{X} = {3\vert\sigma_3(r)\vert + 8 + a_1  - 6\sigma_1(r) - b_1 \over 3}, ~~~~
\mathbb{Y} = {24\sigma_1(r) - 3\vert\sigma_3(r)\vert + 4b_1  - 8 - a_1 \over 3}. \nd
The above set of solutions has one interesting property: it doesn't matter which of the two monotonically increasing functions dominate. To see this, let us first take $\sigma_1(r) = + \vert\sigma_1(r)\vert$. If 
$\vert\sigma_3(r)\vert > \vert\sigma_1(r)\vert$, then $\vert\sigma_3(r)\vert$ still needs to lie between the following range:
\bg\label{rebhall}
{6\vert\sigma_1(r)\vert + b_1  - 8 - a_1 \over 3} ~ < ~ \vert\sigma_3(r)\vert ~ < ~ {24\vert\sigma_1(r)\vert + 4b_1  - 8 -a_1 \over 3},\nd
to keep $\mathbb{X}$ and $\mathbb{Y}$ positive. Now imagine we fix ($a_1, b_1$), and if $b_1  > 8 + a_1$, then the lower bound of $\vert\sigma_3(r)\vert$ in \eqref{rebhall} can be satisfied. However the upper bound is now problematic for large enough $r$, because $\mathbb{Y}$ can become negative unless 
$2\vert\sigma_1(r)\vert < \vert\sigma_3(r)\vert < 8\vert\sigma_1(r)\vert$. For the simplest case where $\sigma_1(r) = 0$, i.e the case where changing $g_s$ automatically changes the $M_p$ scalings, the above restriction is not practical. In the grounds of genericity that we want to impose, all we can say here is that if for large $r$,
$\vert\sigma_3(r)\vert$ dominates over $\vert\sigma_1(r)\vert$, then there are no operators to order 
${g_s^{a_1/3}\over M_p^{b_1}}$. Thus there are only finite number of operators up-to some fixed $r$. This is illustrated in {\bf figure \ref{fig2}}.

For the case with $\vert\sigma_1(r)\vert > \vert\sigma_3(r)\vert$, the story is somewhat similar. To keep both ($\mathbb{X}, \mathbb{Y}$) positive, $\vert\sigma_1(r)\vert$ should lie between the range:
\bg\label{scarjon}
{3\vert\sigma_3(r)\vert + 8 + a_1  - 4b_1 \over 24} ~ < ~ \vert\sigma_1\vert ~ < ~ {3\vert\sigma_3(r)\vert + 8 + a_1  - b_1 \over 6}. \nd
If we take $b_1 < {8 + a_1\over 4}$, then the lower bound is automatic, but the upper bound becomes problematic if we want to keep $\mathbb{X}$ positive for a fixed choice of ($a_1, b_1$). This clearly spells out that the number of operators at any order in ${g_s^{a_1/3}\over M_p^{b_1}}$ can only be finite. See 
{\bf figure \ref{fig2}}.

\begin{figure}[h]
\centering
\begin{tabular}{c}
\includegraphics[width=0.8\textwidth]{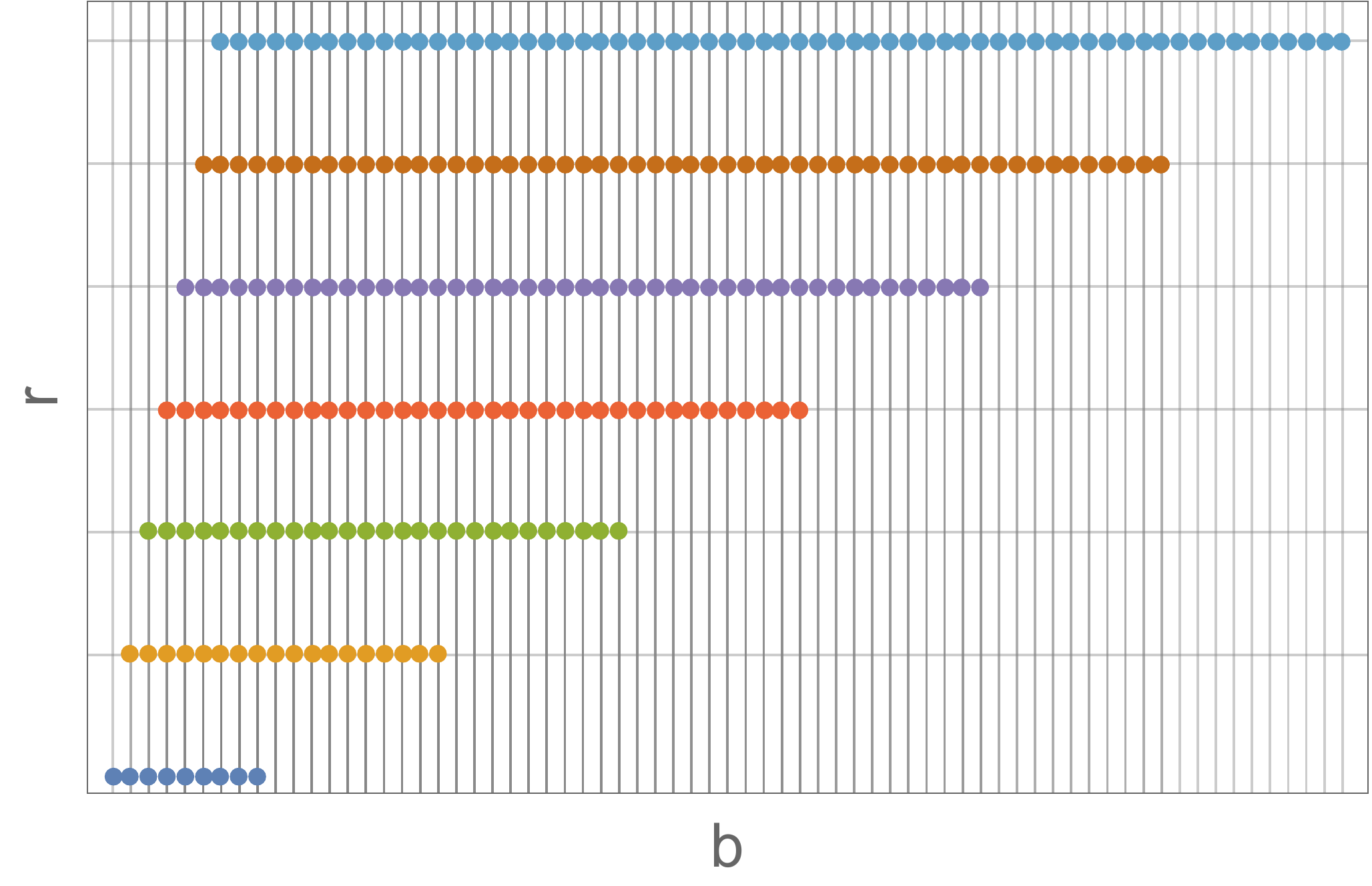}
\end{tabular}
\caption{Again for $\sigma_3(r) \neq {\rm constant}$, the number of operators with the same $g_s$ scaling can be different for different levels of non-locality.  For $\sigma_1(r) < 0, \sigma_3(r) < 0$ there are still finitely many operators with a given $M_p$.}
\label{fig3}
\end{figure}

\begin{figure}[h]
\centering
\begin{tabular}{c}
\includegraphics[width=0.8\textwidth]{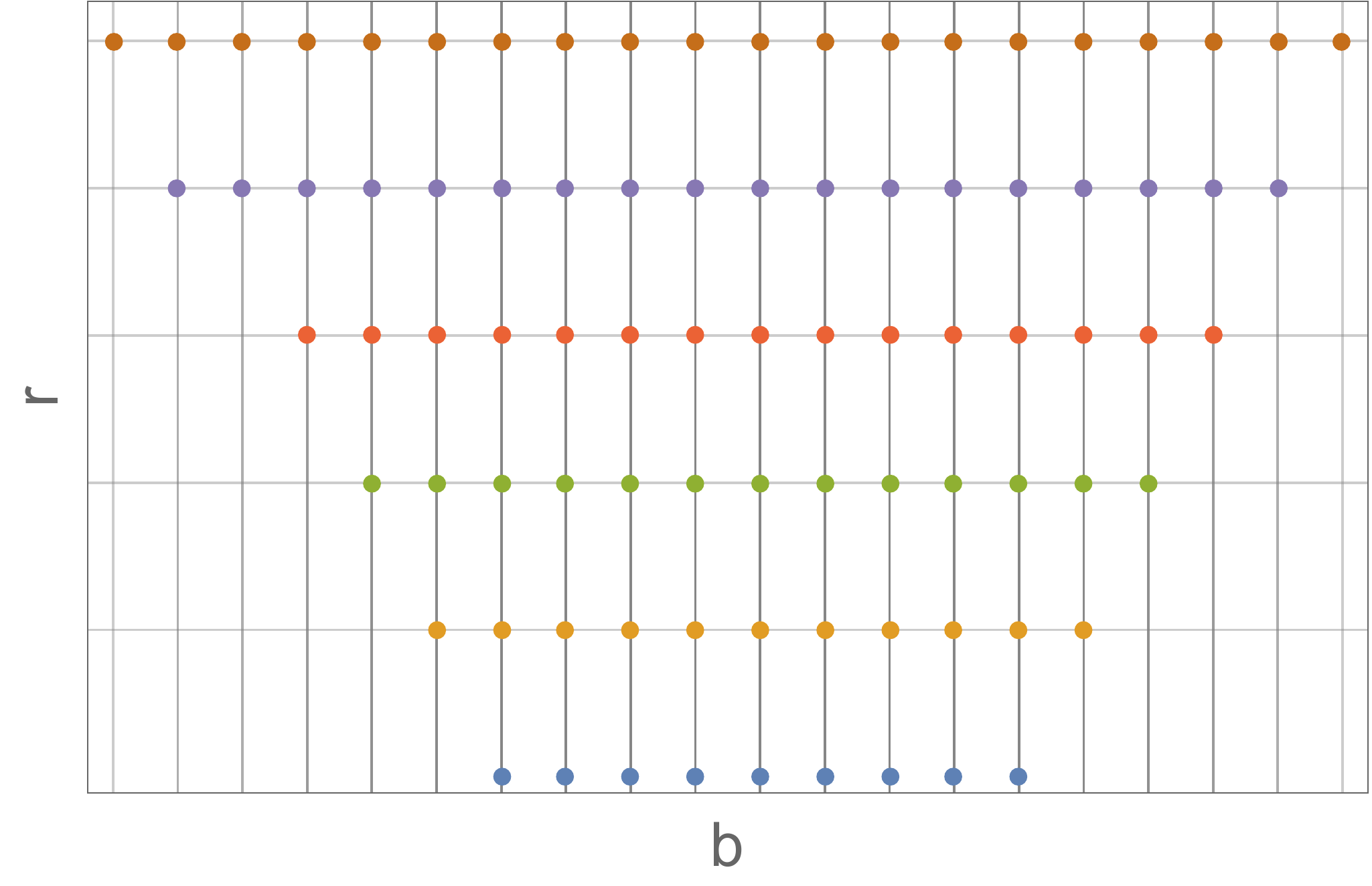}
\end{tabular}
\caption[]{For $\sigma_1(r) > 0$ and falling within the window \eqref{rebhall} for all $r$ the slope of the left boundary of our region of allowed operators changes sign, resulting in an infinite set of operators for each $M_p$ scaling and a breakdown of EFT.}
\label{fig4}
\end{figure}

For the second case when $\sigma_1(r) = -\vert\sigma_1(r)\vert$, we are no longer required to worry about which of the two functions, $\vert\sigma_3(r)\vert$ and $\vert\sigma_1(r)\vert$, dominate as the relative sign between them goes away. This may be seen from the expressions of ($\mathbb{X}, \mathbb{Y}$) that take the following form now:
\bg\label{amason}
\mathbb{X} = {3\vert\sigma_3(r)\vert + 8 + a_1  + 6\vert\sigma_1(r)\vert - b_1 \over 3}, ~~~~
\mathbb{Y} = {4b_1 - 24\vert\sigma_1(r)\vert - 3\vert\sigma_3(r)\vert  - 8 - a_1 \over 3}. \nd
It is now easy to see that, once we fix ($a_1, b_1$), the monotonic increment nature of both 
$\vert\sigma_1(r)\vert$ and $\vert\sigma_3(r)\vert$ would quickly make $\mathbb{Y}$ negative definite. Therefore there cannot be any operators beyond some upper bound of $r$ at any given order in 
${g_s^{a_1/3} \over M_p^{b_1}}$. This is shown in {\bf figure \ref{fig3}}. The special case where $\sigma_3(r)$ falls within the window \eqref{rebhall}, the distribution of the operators is shown 
 in {\bf figure \ref{fig4}}. 

Finally, what happens for the time-independent case, i.e the case when the flux components are time-independent? For the $2+1$ dimensional space-time EOMs, the conditions are now similar to 
\eqref{jhograO}, namely:
\bg\label{fasade}
&&\mathbb{N}_1 + \mathbb{N}_2 + \mathbb{N}_3 ~ = ~ 6\sigma_1(r) + b_1\nonumber\\
&& 4 \mathbb{N}_1 - 2 \mathbb{N}_2 + \mathbb{N}_3 ~ =  -3\sigma_3(r) +  8 + a_1, \nd
and only differ by the presence of the $\sigma_3(r)$ function. The signs of $\sigma_1(r)$ and $\sigma_3(r)$ 
are now irrelevant, and so are their actual behavior with respect to $r$ and with respect to each other. This is because, once we subtract these two set of equations, we get:
\bg\label{machele}
\mathbb{N}_1 - \mathbb{N}_2 = {8 + a_1 - 3\sigma_3(r) - 6\sigma_1(r) -b_1\over 3}, \nd
where, because of the relative {\it minus} sign on the LHS, the RHS can have any signs. Interestingly for any fixed choice for 
($a_1, b_1$), now no longer puts any constraints on the number of operators and therefore the system can have an infinite number of operators at any order in ${g_s^{a_1/3}\over M_p^{b_1}}$. Switching on $n_3$
gives us:
\bg\label{caradev}
2\mathbb{N}_1 - 4\mathbb{N}_2 - \mathbb{N}_3 + 3\sigma_3(r) + 12\sigma_1(r) = 8, \nd
at zeroth order in $g_s$ and $M_p$, similar to what we had in \eqref{amberH}. The relative minus signs of 
$\mathbb{N}_i$, for any sign choices for ($\sigma_1(r), \sigma_3(r)$)
will again ruin an EFT description for the theory in the time-independent case.

\vskip.1in

\noindent {\it Case 3: Neither the $g_s$ nor the $M_p$ scalings change}

\vskip.1in

\noindent This is a special case where to any order in $r$, i.e to any order in non-localities, not only there are finite number of operators for the time-dependent case, their $M_p$ scalings (although would differ when we go from one operator to another for a given $r$) would be {\it similar} when we go from one level of non-locality to another. This is because of the following choice in \eqref{spivibhalo2}:
\bg\label{clobalti}
\sigma_i(r) = \sigma_i(r-1), \nd
where $i = 1, 2, 3$. This means once we know the $g_s$ and the $M_p$ scalings of the local operators, we also know the informations for all the non-local operators. It is not clear whether such a case can be realized in M-theory, but let us nevertheless investigate the consequence of such a choice in \eqref{spivibhalo2}. Clearly now ($\mathbb{X}, \mathbb{Y}$) take the following form:
\bg\label{pasadena}
\mathbb{X} = {8 + a_1 - b_1\over 3}, ~~~~~\mathbb{Y} = {4b_1 - 8 - a_1\over 3}. \nd
It is clear that there are multiple choices for ($a_1, b_1$) which may keep ($\mathbb{X}, \mathbb{Y}$) positive. For example all choices $b_1$ with ${8 + a_1\over 4} < b_1 < 8 + a_1$ are allowed. Does this imply that there are an infinite number of operators to order ${g_s^{a_1/3}\over M_p^{b_1}}$ with $b_1$ in the above range? For example, with $a_1 = 0$, this would imply an infinite number of operators between 
${\cal O}(1/M_p^2)$ and ${\cal O}(1/M_p^8)$. 

The answer to this conundrum is rather simple. Indeed there is an infinite number of operators at any given order in ($a_1, b_1$), but all these operators are simply integrals of the local operator, i.e the operator for $r = 0$, by the function 
$\mathbb{F}^{(r)}(y - y')$ in \eqref{spivibhalo2}. Once we sum up all these operators, the integrations by the non-locality functions simply {\it renormalize} the local operator! In fact all the local operators are renormalized in the {\it same} way because the function $\mathbb{F}^{(r)}$ remains fixed for a given choice of $r$. The possibility of such renormalization should be inherently built-in in the short distance behavior of M-theory itself. 

In the time-independent case, we again expect renormalized local operators, but now the number of such operators is not finite. The EFT breakdown now happens, not due to any non-localities, but due to the 
{\it standard breakdown} that we already discussed in section \ref{local} (see also \cite{maxpaper}).

\section{De Sitter solutions in the landscape \label{later}}

In the above sections we provided detailed arguments to justify why we expect de Sitter solutions to appear in the landscape and not in the swampland, once we switch on time-dependent fluxes and the corresponding quantum corrections. As we showed above, the latter could be very generic, both in terms of local and non-local contributions.  Here we discuss few other related issues concerning the existence of de Sitter solutions in the landscape.

\vskip.1in

\noindent $\bullet$ {\it How do the de Sitter solutions take care of the no-go conditions?}

\vskip.1in

\noindent The no-go conditions as proposed in \cite{giburam}, and updated in \cite{nogo}, may be used to propose an {\it exact} expression for the four-dimensional cosmological constant $\Lambda$ in the IIB side
as shown in \cite{nodS, deft}. This takes the form \cite{deft}:
\bg\label{ambaagun}
\Lambda = {1\over 12 V_6}\left[\mathbb{Q}^i_i - {\mathbb{Q}_{\mathbb{T}^2}\over 2H_0^4} - 
{\mathbb{Q}_{{\cal M}_4 \times {\cal M}_2}\over 4H_0^4} - {2\kappa^2 T_2n_b \over H_0^8} - 
{5\over 32 H_0^8} \Big\langle {\bf G}_{MNab} {\bf G}^{MNab}{\Big\rangle}\right], \nd
where $H_0$ is the constant warp-factor over the internal eight-manifold; $\mathbb{Q}^i_i, \mathbb{Q}_{\mathbb{T}^2}$ and $\mathbb{Q}_{{\cal M}_4 \times {\cal M}_2}$  are the expectation values of the 
{\it quantum} energy-momentum tensors in the spatial, the toroidal ${\mathbb{T}^2\over {\cal G}}$ and 
the base ${\cal M}_4 \times {\cal M}_2$ parts of the Einstein's EOMs as derived in \cite{deft};  $V_6$ is the volume of the base; 
$\kappa, T_2$ and $n_b$ are the data related to the dynamical M2-branes; and the last term in 
\eqref{ambaagun} denotes the expectation value of the G-flux components relevant for the Einstein's EOMs to zeroth order in $g_s$. This has been rigorously derived in \cite{deft}, so we will not go into further details on \eqref{ambaagun} except to point out two crucial facts: one, all of the terms in \eqref{ambaagun}, except the first one, appear with minus signs; and two, in the time-independent case, the quantum pieces are constructed out of an infinite number of contributions with no $g_s$ or $M_p$ hierarchies as we showed earlier. This means the series representation of $\Lambda$ cannot be terminated and therefore \eqref{ambaagun} has no solution in the time-independent case (see also \cite{shakla} for the IIA case). Once we switch on time-dependences, all the quantum pieces have finite number of contributions. 
$\mathbb{Q}^i_i$ can then be made to dominate over all other terms, as the other two quantum pieces simply renormalize the last term without changing its overall sign and in-turn keeping it small\footnote{Recall that all our flux components are kept very small, although their derivatives may become of order $M_p$ as we saw in \eqref{canyon} and \eqref{alexmey}. Also contributions to $\mathbb{Q}^i_i$ are from 
the eight derivative terms, whereas contributions to $\mathbb{Q}_{\mathbb{T}^2}$ and 
 $\mathbb{Q}_{{\cal M}_4 \times {\cal M}_2}$ only come from the two derivative terms in \eqref{phingsha0}.}
\cite{deft}, thus realizing $\Lambda > 0$ solution in the IIB side.

\vskip.1in

\noindent $\bullet$ {\it How do the de Sitter solutions remain outside the swampland?} 
 
\vskip.1in

\noindent The original swampland criteria, as defined in \cite{vafa}, basically deals with an inequality of the form $\vert\nabla V\vert/ V > c$, where $c = {\cal O}(1)$ number. The claim is that if the ratio falls below the prescribed value of $c$, the theory should have no UV completion and should thus be discarded to the swampland. Computing the type IIB potential requires duality chasing all the quantum terms. A slightly different but simpler alternative is to look at the potential of the $2+1$ dimensional theory on the M-theory side. This too should satisfy the gradient criterion. For this case the full quantum potential \eqref{ducksoup2} {\it can} be computed in the time-dependent cases, as we rigorously demonstrated above. This implies, a dimensional reduction over the compact internal space will convert all the fluctuations of the G-flux and metric components into scalar fluctuations $\varphi^{(i)}$, thus reproducing:
\bg\label{linaolina}
{\left\vert\nabla \mathbb{V}_Q \right\vert \over \mathbb{V}_Q} = \sum_{i, j}^{{\rm dim}\left({\cal M}_\varphi\right)} {\sqrt{g^{\varphi^{(i)} \varphi^{(j)}}
\partial_{\varphi^{(i)}} \mathbb{V}_Q \partial_{\varphi^{(i)}}\mathbb{V}_Q} \over \mathbb{V}_Q}  = 
{\cal O}\left(\sum_{k= 1}^{{\rm dim}\left({\cal M}_\varphi\right)} {1\over g_s^{n_k}}\right) >> 1, \nd    
as derived in \cite{deft} with $g_{\varphi^{(i)} \varphi^{(j)}}$ being the metric on the moduli space 
${\cal M}_\varphi$ of the scalar fields  $\varphi^{(i)}$. The scalar fields are taken as canonical, meaning that all $g_s$ (and other spatial) dependences are absorbed in their definitions in such a way that the metric on the moduli space is flat. Note also that these scalars are {\it fluctuations} over our de Sitter background and therefore their $g_s$ dependences are not necessarily inherited from the corresponding background values. 
Clearly if all the G-flux components are 
time-dependent, the RHS of \eqref{linaolina} is always very large in the $g_s \to 0$ limit. 
However, even if we keep some of the G-flux components time-independent (i.e $k_1 \ge 0$ and 
$k_2  \ge {3\over 2}$ in \eqref{melamon0}), the sum over the moduli space in \eqref{linaolina} will again guarantee that the swampland bound is easily overcome. Such a conclusion is not surprising in the light of what we discussed in sections \ref{qotom1} and \ref{qotom2} namely, once time-dependences are switched on, natural $g_s$ and $M_p$  hierarchies allow four-dimensional EFT description to appear in the IIB side.

\vskip.1in

\noindent $\bullet$ {\it What about Newton's constant, moduli stabilization and fermions?}

\vskip.1in

\noindent In a time-dependent internal space of the form \eqref{pyncmey2} one would generically argue that the Newton's constant no longer remains a {\it constant}, but becomes a time-dependent function. However, this is not always true, and in particular if we can maintain $F_1(t) F_2^2(t) = 1$ under full quantum corrections, the Newton's constant will remain truly a constant. One of the results of \cite{deft} is the rigorous derivation that this is indeed possible. In the late-time limit none of the internal subspaces, namely 
${\cal M}_4$, ${\cal M}_2$ and ${\mathbb{T}^2\over {\cal G}}$, become singular and therefore in the IIB side we expect a four-dimensional space with de Sitter isometries, and a time-independent Newton's constant to appear under the full effects of the local and the non-local quantum corrections. Of course this would make sense if the moduli are properly stabilized, so the pertinent question is: how
are the moduli stabilized in such compactifications? One would however first need to understand what moduli {\it stabilization} mean in a background that is changing dynamically. The answer is not hard to see: the moduli should be stabilized at every {\it instant} by fluxes and the quantum corrections. Once the fluxes and the quantum corrections change with time, the moduli should respond accordingly. To see this, let us consider the complex structure moduli. The time-independent parts of the G-flux components would make sure that they are stabilized even in the late time limit where $g_s \to 0$. After which the $g_s$ dependent terms in say \eqref{frostgiant2} would in turn make sure that there is a proper dynamical evolution of the moduli without any Dine-Seiberg runaway. On the other hand, fixing the radial modulus (or the K\"ahler moduli) typically require fermionic condensates \cite{DRS1, kachruS}. So the question is how the fermions appear in our analysis. The answer is again not hard to see from the G-flux ansatze \eqref{canyon}, by considering only the first term. As discussed in \cite{DEMf}, one may express the anti-symmetric G-flux components by adding a fermionic completion of the form:
\bg\label{libyasha}
{\bf G}_{MNab}(x, y, y^a) = \bar{\Psi} \Gamma_{MN} \Psi(x, y) \otimes \Omega_{ab}(y^a), \nd
where $x^\mu$ denotes the coordinates of $2+1$ dimensional space-time, $y = (y^m, y^\alpha)$ denotes the coordinates of ${\cal M}_4 \times {\cal M}_2$; $y^a$ denotes the coordinates of  
${\mathbb{T}^2\over {\cal G}}$;  $\Gamma_{MN}$ is the anti-symmetric product of nine-dimensional gamma matrices;
and $\Omega_{ab}$ denotes the two-form expressed generically as 
\eqref{alexmey} with an unwarped metric in the time-independent case and with a warped metric in the time-dependent case. 
One can then further sub-divide $\Psi(x, y) = \psi(x) \otimes \chi^{(6)}(y)$ where $\chi^{(6)}(y)$ being the fermion component in the internal space and $\psi(x)$ is the space-time fermion component. Plugging the G-flux 
\eqref{libyasha} in \eqref{phingsha0} will lead to powers of condensates of the form 
$\left(\bar{\psi} \psi\right)^n$ that would contribute to the quantum terms. In the time-independent case, as we saw in sections \ref{qotom1} and \ref{qotom2}, there are infinite number of such terms to zeroth order in $g_s$ and $M_p$, thus ruining the four-dimensional EFT description in the system. Only in the time-dependent case, an EFT description is possible.

\vskip.1in

\noindent $\bullet$ {\it How are the flux quantization, anomaly cancellation and EOMs achieved?}

\vskip.1in

\noindent As we know from \cite{wittenflux, BB1, DRS1}, switching on G-flux components on a compact four-manifold leads to issues like flux quantization and anomaly cancellation. These have already been resolved for the time-independent case, so the question is what happens when both the fluxes and the internal manifold become time-dependent. In \cite{deft} we provided a detailed derivations of the flux quantization scheme, anomaly cancellation conditions and the supergravity EOMs in the presence of quantum terms. Since in this paper we only discuss computations {\it not} performed in \cite{deft}, we refer the readers to 
section 4 of \cite{deft} for details regarding the above questions. We do however want to point out two things. One, the hierarchical set of quantum terms, at every order in $g_s$, is responsible for keeping the metric components at zeroth order unchanged so that \eqref{pyncmey2} can give rise to an exact de Sitter solutions in the IIB side. This delicate balancing act is important and in \cite{deft} we carefully showed how this may be achieved for our case in the time-dependent case. In fact the finiteness of the number of quantum terms at every order in $g_s$ served as the key ingredients there.  Two, in the quantum potential 
$\mathbb{V}_Q$, only a {\it part} contributes to the cosmological constant $\Lambda$ \eqref{ambaagun}. 
In fact this contribution is from the zeroth order in $g_s$. Going to the higher order in $g_s$, switches on G-flux components as may be seen from \eqref{frostgiant2} which, in turn, back-react on the background creating negative gravitational potentials. These negative contributions are nullified by the positive contributions from the higher order $g_s$ terms in \eqref{ducksoup2}. This way the cosmological constant in 
\eqref{ambaagun} gets no additional contributions from the higher order $g_s$ dependent pieces of 
\eqref{ducksoup2}.  Combined with the time-independent Newton's constant discussed earlier, this results 
in a genuine four-dimensional de Sitter solution in the IIB side.

\section{Discussions and conclusions}

In this paper we hopefully provided convincing arguments to justify the presence of a de Sitter solution \eqref{pyncmey2} in the IIB string {\it landscape} and {not} in the {\it swampland}. The de Sitter solution, that survives a generic choice of local and non-local quantum terms, appears with a renormalized cosmological constant  
$\Lambda$ \eqref{ambaagun} and Newton's constant that are time-independent. 
Our strategy here is to provide explicit criteria of the existence and the non-existence of EFT descriptions in the time-dependent and the time-independent cases respectively. In fact the impossibility of an EFT description in the time-independent case provides some credence to the swampland idea \cite{vafa} but, as demonstrated here, these swampland criteria are not restrictive enough to forbid four-dimensional de Sitter solutions to exist in the time-dependent case, at least the late time ones\footnote{Our choice of the normalizable two-form \eqref{alexmey} might even suggest that time-dependent solutions  are more natural than the time-independent ones.}. 
 
Question however is, what happens for the {\it early} times i.e for $\Lambda \vert t\vert^2 \to 1$ in our language? We expect early time physics to be an inflationary one, and there has been some recent works \cite{vahid} to argue how certain inflationary models, like warm inflation, may be outside the swampland. However the way we have set-up our analysis, while providing a consistent picture in the $t \to 0$ or $g_s \to 0$, doesn't quite generalize to the $g_s \to 1$ limit where the system becomes strongly coupled. The latter limit requires a more intricate construction because even a duality that takes $g_s \to {1\over g_s}$, doesn't quite help\footnote{For example, now we need to worry about ${\cal O}\left(e^{-1/g_s}\right)$ contributions from wrapped branes and instantons in addition to the perturbative terms from 
\eqref{phingsha0}.}. 
It does however succinctly map the late-time physics to the big-bang time (see footnote 
\ref{godfather}). We will leave this interesting question for future publication, but here we can at least  make  one observation. The M2-branes appearing in our set-up, with $n_b > 0$ in \eqref{ambaagun}, naturally give rise to a D3-D7 system in IIB, where the seven-branes come from the orbifold singularities in the internal space. This means at early times we might be able to study a D3-D7 inflationary model, much along the lines of \cite{DHHK} (see \cite{ding} for the possibility of realizing other models of string cosmology).

Yet another question is the Wilsonian renormalization or the Wilsonian effective action at any given energy scale. As discussed earlier (see also footnote \ref{tcc}) this is subtle not only because of the red-shifting UV modes 
\cite{transplanckian}, 
but also because of the presence of the winding modes from the M2 and M5-branes that behave differently from the momentum modes in an accelerating background. The precise procedure of {\it integrating out} the heavy modes is non-trivial and has not been attempted in the literature yet (see some discussion on this in \cite{xue}). The subtlety lies in the proper treatment of the UV modes in string and M theories as we know that neither of these theories are in the swampland, i.e both these theories have well defined UV behavior. Question then is whether the trans-Planckian modes lead to non-unitary behavior when de Sitter solutions are chosen. Our analysis here showed us that for generic choice of local \eqref{phingsha0}, and non-local 
\eqref{karalura2}, quantum terms there is a well defined de Sitter solution once the internal degrees of freedom become time-dependent. Does TCC pose an issue for such a scenario? This is a pertinent
question that requires a careful investigation\footnote{Let us imagine that TCC does pose an issue for such a background. It would then imply that the time-dependent background \eqref{pyncmey2} should be in the swampland, i.e should have no UV completion, contradicting the very fact that we get a {\it finite} number of operators at every scale. One way out of this conundrum will be to allow for a well-defined UV dynamics in string theory with a background like \eqref{pyncmey2}. If not, the puzzle is more severe here.}. 
Additionally there is also the issue with the un-cancelled vacuum energies from all these modes that desperately await proper regularization and renormalization schemes \cite{danielsson2}. As hinted here and also in \cite{deft}, such insurmountable problems might have an easier solution if we instead view the de Sitter background in IIB (or its M-theory uplift) as a coherent or a squeezed-coherent state over a supersymmetric solitonic background. In the latter procedure the modes are well defined and the vacuum energies automatically cancel, so a Wilsonian effective action may be defined easily. The de Sitter solution then appears as the most probable configuration in the Hilbert space. 
To decide which of these two viewpoints would eventually prevail is an important question and deserves further effort.


\section*{Acknowledgements}

 We would like to thank Suddhasattwa Brahma and Robert Brandenberger for many helpful discussions and comments on the draft;  and  Cumrun Vafa and Edward Witten for useful correspondences.
The work of KD, ME and MMF is supported in part by the Natural Sciences and Engineering Research Council of Canada.


\end{document}